\documentclass[twocolumn]{aastex61}
\pdfoutput=1 
\usepackage{amsmath,amstext}
\usepackage[T1]{fontenc}
\usepackage{apjfonts} 
\usepackage[figure,figure*]{hypcap}

\shortauthors{Lockhart et al.}

\begin{document}

\title{Characterizing and Improving the Data Reduction Pipeline for the Keck OSIRIS Integral Field Spectrograph}

\author[0000-0002-8130-1440]{Kelly E. Lockhart}
\affiliation{Harvard-Smithsonian Center for Astrophysics, 60 Garden Street, Cambridge, MA 02138}
\affiliation{Institute for Astronomy, University of Hawaii, Manoa, Honolulu, HI 96822}
\email{kelly.lockhart@cfa.harvard.edu}
\author[0000-0001-9554-6062]{Tuan Do}
\affiliation{Department of Physics \& Astronomy, University of California, Los Angeles, CA, USA 90095}
\author[0000-0001-7687-3965]{James E. Larkin}
\affiliation{Department of Physics \& Astronomy, University of California, Los Angeles, CA, USA 90095}
\author[0000-0003-0439-7634]{Anna Boehle}
\affiliation{Institute for Particle Physics and Astrophysics, ETH Zurich, CH-8093 Zurich, Switzerland}
\affiliation{Department of Physics \& Astronomy, University of California, Los Angeles, CA, USA 90095}
\author[0000-0002-3289-5203]{Randy D. Campbell}
\affiliation{W. M. Keck Observatory, Waimea, HI, USA 96743}
\author{Samantha Chappell}
\affiliation{Department of Physics \& Astronomy, University of California, Los Angeles, CA, USA 90095}
\author[0000-0003-3765-8001]{Devin Chu}
\affiliation{Department of Physics \& Astronomy, University of California, Los Angeles, CA, USA 90095}
\author{Anna Ciurlo}
\affiliation{Department of Physics \& Astronomy, University of California, Los Angeles, CA, USA 90095}
\author{Maren Cosens}
\affiliation{Center for Astrophysics \& Space Sciences, University of California San Diego, 9500 Gilman Dr, CA, 92039 USA}
\affiliation{Department of Physics, University of California San Diego, 
9500 Gilman Drive, La Jolla, CA 92093 USA}
\author[0000-0002-0176-8973]{Michael P. Fitzgerald}
\affiliation{Department of Physics \& Astronomy, University of California, Los Angeles, CA, USA 90095}
\author[0000-0003-3230-5055]{Andrea Ghez}
\affiliation{Department of Physics \& Astronomy, University of California, Los Angeles, CA, USA 90095}
\author[0000-0001-9611-0009]{Jessica R. Lu}
\affiliation{Astronomy Department, University of California, Berkeley, CA, USA, 94720}
\author{Jim E. Lyke}
\affiliation{W. M. Keck Observatory, Waimea, HI, USA 96743}
\author{Etsuko Mieda}
\affiliation{NRC Herzberg Astronomy and Astrophysics, 5071 West Saanich Rd, Victoria, BC, V9E 2E7, Canada}
\author{Alexander R. Rudy}
\affiliation{Department of Astronomy, University of California, Santa Cruz, CA 95064}
\author[0000-0002-0710-3729]{Andrey Vayner}
\affiliation{Center for Astrophysics \& Space Sciences, University of California San Diego, 9500 Gilman Dr, CA, 92039 USA}
\affiliation{Department of Physics, University of California San Diego, 
9500 Gilman Drive, La Jolla, CA 92093 USA}
\author{Gregory Walth}
\affiliation{Center for Astrophysics \& Space Sciences, University of California San Diego, 9500 Gilman Dr, CA, 92039 USA}
\author{Shelley A. Wright}
\affiliation{Center for Astrophysics \& Space Sciences, University of California San Diego, 9500 Gilman Dr, CA, 92039 USA}
\affiliation{Department of Physics, University of California San Diego, 
9500 Gilman Drive, La Jolla, CA 92093 USA}

\begin{abstract}
OSIRIS is a near-infrared (1.0--2.4 $\micron$) integral field spectrograph operating behind the adaptive optics system at Keck Observatory, and is one of the first lenslet-based integral field spectrographs. Since its commissioning in 2005, it has been a productive instrument, producing nearly half the laser guide star adaptive optics (LGS AO) papers on Keck. The complexity of its raw data format necessitated a custom data reduction pipeline (DRP) delivered with the instrument in order to iteratively assign flux in overlapping spectra to the proper spatial and spectral locations in a data cube. Other than bug fixes and updates required for hardware upgrades, the bulk of the DRP has not been updated since initial instrument commissioning. We report on the first major comprehensive characterization of the DRP using on-sky and calibration data. We also detail improvements to the DRP including characterization of the flux assignment algorithm; exploration of spatial rippling in the reduced data cubes; and improvements to several calibration files, including the rectification matrix, the bad pixel mask, and the wavelength solution. We present lessons learned from over a decade of OSIRIS data reduction that are relevant to the next generation of integral field spectrograph hardware and data reduction software design. 

\end{abstract}


\section{Introduction}
\begin{figure}
\begin{center}
\includegraphics[width=\columnwidth]{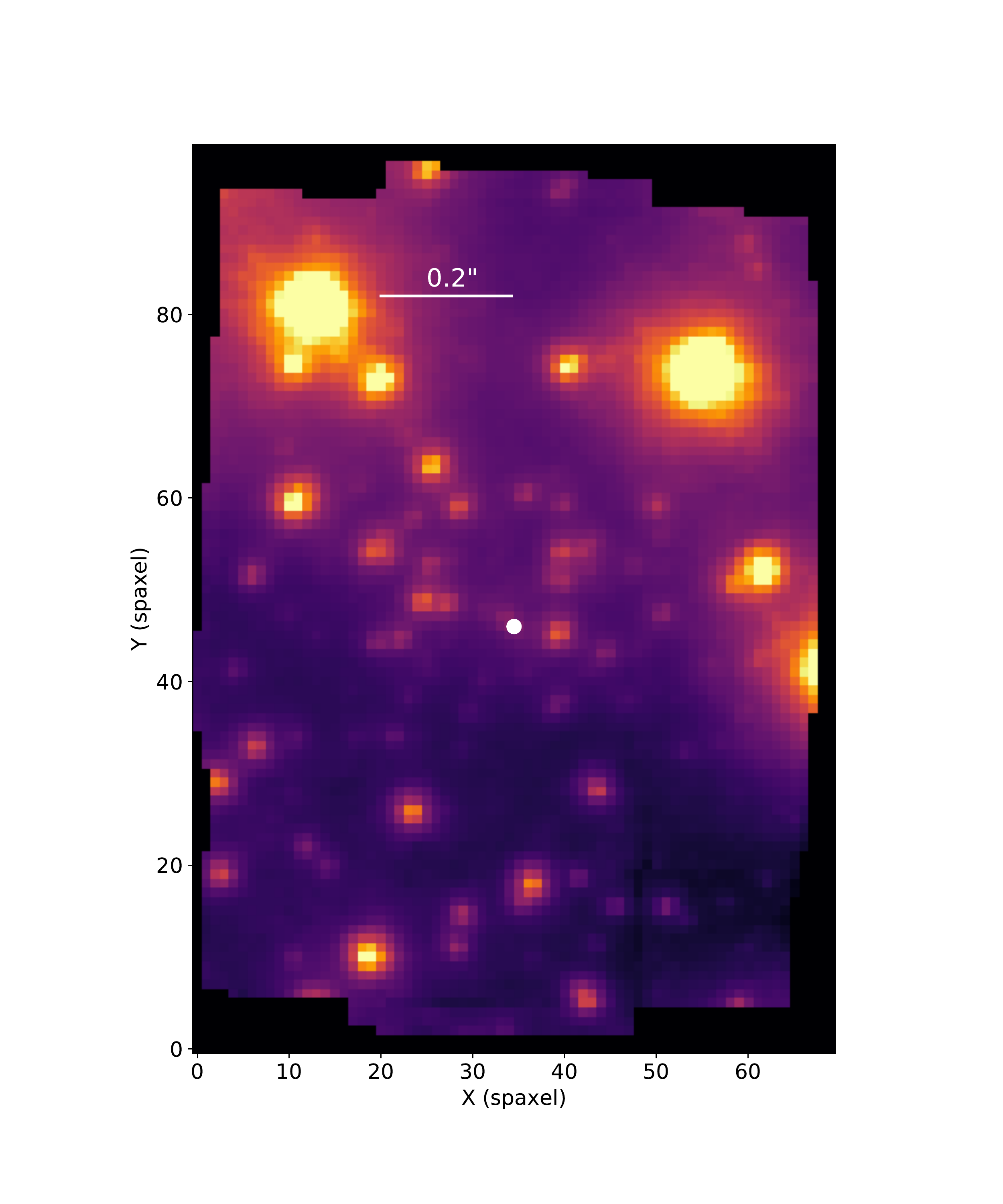}
\caption{An OSIRIS observation of the Galactic center at the 35 mas plate-scale in the Kn3 filter. The NIR emission from the supermassive black hole, Sgr A*, is labeled with a white dot. The data are taken at a position angle of 175 degrees. The observations take advantage of the Keck LGS AO system.}
\label{fig:gc}
\end{center}
\end{figure}

The OSIRIS \citep[OH Suppressing InfraRed Imaging Spectrograph;][]{larkin2006osiris:} instrument behind the Adaptive Optics (AO) system at the
W.~M.~Keck Observatory has proven to be an important tool for
extragalactic, Galactic, and solar system observational
astrophysics. OSIRIS has produced nearly one-half of the laser guide
star (LGS) AO publications from Keck since its commissioning in 2005,
including the majority of the extragalactic LGS papers. Results from
OSIRIS observations span fields from planetary science \citep[e.g.][]{laver2009the-global, laver2009component-resolved, brown2013salts}, some of the first spectroscopic
characterization of extrasolar planets,
\citep[e.g.][]{bowler2010near-infrared, barman2011clouds,
konopacky2013detection}, studies of crowded fields such as the
center of the Milky Way galaxy \citep[e.g.][]{do2013three-dimensional,lu2013stellar,yelda2014properties,lockhart2018}, measurement of
the masses of supermassive black holes (SMBHs) in nearby galaxies
\citep[e.g.][]{mcconnell2011two-ten-billion-solar-mass,medling2011mass,walsh2012a-stellar}, and studies of
high redshift (1 < z < 3) galaxies \citep[e.g.][]{wright2007integral,stark2008the-formation,law2009the-kiloparsec-scale,jones2010resolved}. 
A sample observation is shown in figure~\ref{fig:gc}, demonstrating OSIRIS's near diffraction-limited spatial resolution.

\begin{figure}
\begin{center}
\includegraphics[width=\columnwidth]{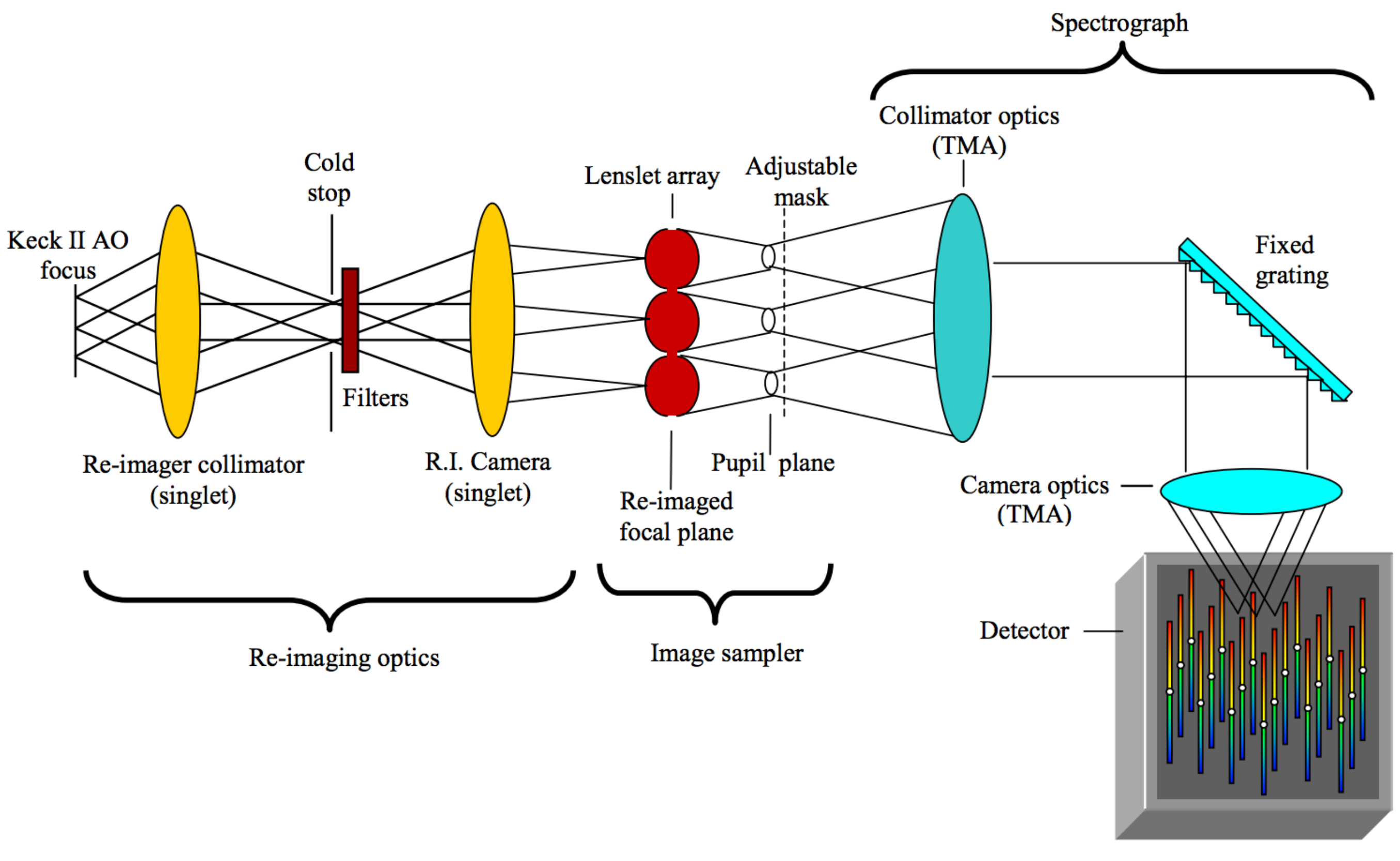}
\caption{Optical configuration of the OSIRIS spectrograph, from the instrument manual, available from the DRP Github page.\textsuperscript{a}}
\small\textsuperscript{a} https://github.com/Keck-DataReductionPipelines/OsirisDRP
\label{fig:optical}
\end{center}
\end{figure}

OSIRIS is an innovator among near-infrared (NIR, 1.0--2.4 $\micron$) integral field
spectrographs (IFS) due to its unique use of a lenslet array to split
the field into spatial pixels (spaxels) while preserving the high Strehl ratio point
spread function (PSF) of the Keck AO system (figure~\ref{fig:optical}). 
The new planet finding spectrographs (e.g., Gemini Planet Imager,
SPHERE) have a similar configuration but with a very different output
data format. 
IFSs are also being planned for the next-generation extremely large telescopes \citep[e.g.][]{2016SPIE.9908E..1WL, 2016SPIE.9908E..1YS, 2016SPIE.9908E..1XT, 2016SPIE.9913E..4AW} and forethought in the data reduction pipeline and necessary calibrations is essential for their success. 

\begin{figure}
\begin{center}
\includegraphics[width=\columnwidth]{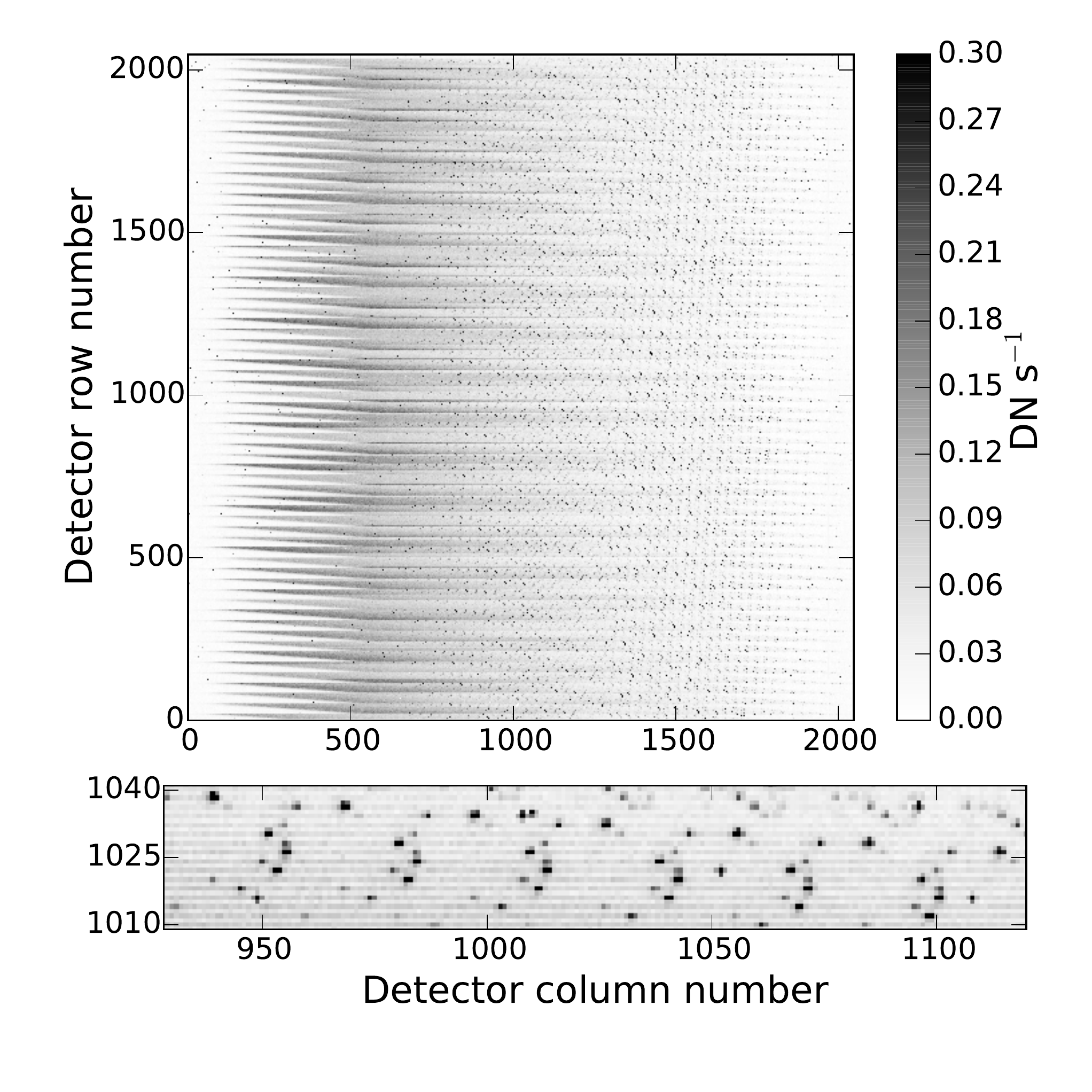}
\caption{The OSIRIS spectra are closely packed on the detector. \emph{Top:} Unreduced 2D detector Kbb/35 sky frame. The wavelength direction increases to the left; the bright bands at the left are the rising thermal background in the red. \emph{Bottom:} Zoom in of the panel above. Individual gray rows, alternating with the fainter background, represent the thermal continuum in individual sky spaxels. Bright spots are OH sky lines. Neighboring spectra are staggered by 32 detector columns, which translates to a stagger of 32 spectral channels in the reduced cube.}
\label{fig:rawdet}
\end{center}
\end{figure}

The OSIRIS IFS was designed to have very stable spectra
that fall at the same position on the detector, using a
single, fixed grating and no moving components after the lenslet
array. In order to maximize the field of view, the spectra from
adjacent spaxels are only separated by 2 pixels (see figure~\ref{fig:rawdet}). In the current configuration, the full-width half-maximum (FWHM) of individual spectra perpendicular to the dispersion direction range from roughly 1.3 to 1.7 pixels \citep{boehle2016upgrade} and thus overlap each other in their wings (though the FWHM was larger and the spectra have overlapped more in previous instrument hardware configurations, as discussed further throughout this work). However,
the stable format enables the use of a deconvolution process in the OSIRIS data reduction pipeline (DRP) to
separate the interleaved spectra in the input 2D detector format and to assign flux to each
spaxel at every wavelength in the output data cube (dimensions x, y, lambda). The DRP is essential for both real-time processing at the telescope and for
post-processing of all data. 

\subsection{The OSIRIS data reduction pipeline}
\label{ssec:drp} 
The complexity of the raw data output from the instrument necessitated a well-developed data reduction pipeline from delivery \citep{krabbe2002data, 2017ascl.soft10021L}, the first for an instrument on Keck. The DRP, written primarily in IDL with computationally intensive processes passed into C, is modularized, such that individual modules of the reduction can be turned on or off as needed at runtime. Typical reduction modules, such as dark subtraction, cosmic ray removal, and telluric correction, are available. Also included is an implementation of the scaled sky subtraction algorithm detailed by \citet{davies2007a-method}. 

The unique part of the DRP is the spatial rectification module (\texttt{spatrectif}), which separates the overlapping spectra in the 2D frame and places flux into corresponding spaxels in the output data cube. The OSIRIS spectral format allocates only 2 pixels between neighboring spectra, and there is a stagger in wavelength of about 32 pixels between lenslet neighbors. While the PSF of each lenslet in the spatial direction, perpendicular to the dispersion direction, is currently below 2 pixels in FWHM, this has not always been true, and in all cases some blending occurs which must be extracted with the deconvolution. The spatial rectification module uses empirically determined rectification matrices (\S~\ref{ssec:recmat}) to map the correspondence between pixels in the 2D frame and spaxels in the data cube. Using the rectification matrices as an initial estimate of the correspondence between pixels in the 2D detector and spaxels in the data cube, the Gauss-Seidel method is used to iteratively assign flux from the cleaned 2D image into individual spaxels. The algorithm is further described by \citet{krabbe2004data} and in the instrument manual\footnote{https://github.com/Keck-DataReductionPipelines/OsirisDRP/blob/master/OSIRIS\_Manual\_v4.2.pdf}. In addition to mapping the lenslet responses, the rectification matrices also serve as flat fields to determine the detector pixel response.

To run the pipeline, the user designates which modules will be used, along with setting any module-specific options or keywords, in an XML file. This XML file is saved to a queue folder that is regularly checked by the DRP. The DRP then processes any new XML files and saves the output as requested.

\subsection{Hardware upgrades}

\begin{figure*}
\begin{center}
\includegraphics[width=6in]{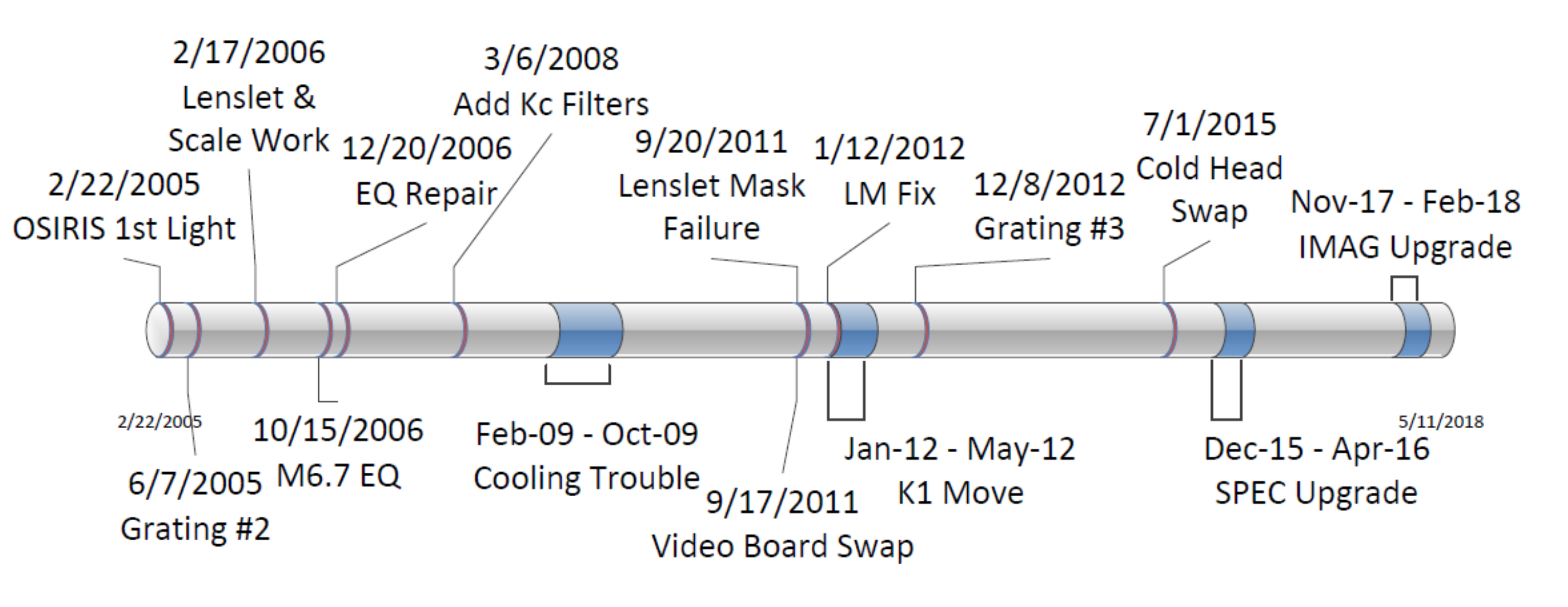}
\caption{Timeline of OSIRIS hardware changes and upgrades since first light.}
\label{fig:timeline}
\end{center}
\end{figure*}

OSIRIS has been operational on Keck for over a decade. Over the course of that that time, it has been subject to repairs and upgrades which are summarized in figure~\ref{fig:timeline}. Two major hardware upgrades have occurred in recent years. 

In December 2012, the dispersion grating was upgraded \citep{mieda2014efficiency}, increasing the grating efficiency and flux on uncleaned raw 2D detector frames by a nearly a factor of 2.  At the same time, OSIRIS was moved from Keck-II to Keck-I, to take advantage of Keck-I's new LGS AO system. At this time, the DRP was updated to include an updated wavelength solution, to account for the different image orientation, to update the world coordinate system (WCS), and to account for differential dispersion between the Keck-II and Keck-I AO system dichroics.

In January 2016, the detector behind the spectrograph portion of OSIRIS was upgraded from a Hawaii-2 detector to a Hawaii-2RG detector \citep{boehle2016upgrade}. The new detector shows a lower dark current and improved throughput by up to a factor of 2 in raw 2D detector frames, due to a combination of an improved quantum efficiency and a reduction in noise sources such as crosstalk between readout channels \citep{boehle2016upgrade}. The DRP was updated with a new wavelength solution, revised locations of the spectra on the 2D frames, and the removal of glitch modules only needed for the old Hawaii-2 detector.

\subsection{Motivation}

Beginning around the same time of the OSIRIS move to Keck-I and the grating upgrade in 2012, observers began to notice irregularities in their reduced data cubes: the presence of unphysically large positive and negative spikes at a few wavelength channels in a small number of spaxels in reduced data, poor performance by the cosmic ray module in successfully identifying and removing cosmic rays, and offsets in the wavelength solution. In addition, observers noticed that reduced science data of QSO observations incorrectly showed dark spaxels near the central bright compact source. Similarly, observations of bright narrow emission lines incorrectly showed absorption or negative flux at the wings of the narrow emission line. Finally, single channel maps of OH sky lines or other evenly illuminated sources were no longer flat, but instead displayed a rippling pattern unrelated to incident illumination. These issues led to an effort to better characterize the behavior of the DRP and to improve the quality of the reduced data.

Since initial delivery, the DRP has historically been maintained by Keck Observatory staff, with coding and testing performed by volunteered time from a few instrument team members. Upgrades to the software have primarily occurred after hardware upgrades that necessitated changes to the wavelength calibration, for instance. However, the scope of the necessary DRP characterization to resolve these issue was beyond the scope of usual DRP maintenance. The OSIRIS Pipeline Working Group was formed and undertook the first major comprehensive characterization of the DRP using calibration and on-sky data. We detail our efforts in the following report. 

We outline the data we used for this work in \S~\ref{sec:data}, artifacts from the flux assignment algorithm in \S~\ref{sec:fluxassign}, and spatial rippling in \S~\ref{sec:spatrip}. We describe improvements to the following calibration files in three sections: the rectification matrices (\S~\ref{sec:crinrecmat}), the bad pixel mask (\S~\ref{sec:badpix}), and the wavelength solution (\S~\ref{sec:wavesol}). Many of the issues and improvements we outline here are relevant to the future generation of IFSs on the next generation of telescopes; we outline our recommendations in \S~\ref{sec:rec}.

The DRP is open-source and is hosted on Github\footnote{https://github.com/Keck-DataReductionPipelines/OsirisDRP} as of May 2016. Releases are made as needed, or roughly once per year. This work uses release v.4.0.0, and many of the improvements detailed here are included in v.4.1.0.

\section{DATA}
\label{sec:data}
Throughout this paper, we will use several on-sky science and engineering data sets to investigate the OSIRIS spectral extraction routine and performance. These data sets are described below and summarized in Table~\ref{tab:obs}.

\begin{deluxetable}{llrcr}
\renewcommand{\arraystretch}{0.9}
\tablecaption{Keck OSIRIS Observations}
\tablewidth{0pt}
\tablehead{
\colhead{Observation} & \colhead{Date} & \colhead{Exposure} & \colhead{Filter} & \colhead{Plate}  \\[-2pt]
\colhead{Type} & \colhead{} & \colhead{Time} & \colhead{} & \colhead{Scale)}  \\[-2pt]
\colhead{} & \colhead{(UT)} & \colhead{(s)} & \colhead{} & \colhead{(mas)} 
}
\startdata
arc & 2012 Jun 15 & 60 & Hbb & 20 \\ 
arc$^a$ & 2013 Apr 11 & 60 & Kbb & 50 \\ 
arc$^a$ & 2015 Apr 06 & 10 & Kbb & 50 \\ 
arc & 2015 Dec 17 & 32 & Kbb & 35 \\ 
arc & 2015 Dec 17 & 16 & Kbb & 50 \\ 
arc & 2015 Dec 17 & 4 & Kbb & 100 \\ 
arc & 2016 Mar 18 & 60 & Jbb & 50 \\ 
arc$^a$ & 2016 Sep 02 & 30 & Kbb & 20 \\ 
arc$^a$ & 2016 Sep 02 & 30 & Kbb & 35 \\ 
arc$^a$ & 2016 Sep 02 & 30 & Kbb & 50 \\ 
arc$^a$ & 2016 Sep 02 & 1.5 & Kbb & 100 \\ 
QSO & 2014 May 19  & 4x600 &Hn3 &100 \\ 
QSO & 2015 Aug 09  & 7x600 & Kn1&100 \\ 
sky & 2012 Jun 09 & 900 & Kn3 & 20 \\ 
sky & 2012 Jun 09 & 900 & Kn3 & 35 \\ 
sky & 2013 May 14 & 900 & Kn3 & 35 \\ 
sky & 2013 May 14 & 900 & Kn3 & 50 \\ 
sky & 2013 May 11 & 900 & Kbb & 35 \\ 
sky & 2014 May 17  & 900 & Kbb & 50 \\ 
sky & 2015 Jul 19 & 100 & Kn3 & 100 \\ 
sky & 2015 Jul 22 & 900 & Hbb & 50 \\ 
sky & 2016 Mar 21 & 600 & Jbb & 50 \\ 
sky & 2016 Mar 21 & 600 & Hbb & 50 \\ 
sky & 2016 Apr 18 & 900 & Jn2 & 35 \\ 
sky & 2016 May 14 & 900 & Kbb & 35\\ 
sky & 2016 Jul 11 & 900 & Kbb & 35 \\ 
sky$^a$ & 2016 Sep 02 & 600 & Kbb & 35 \\ 
sky$^a$ & 2016 Sep 02 & 600 & Kbb & 50 \\ 
sky$^a$ & 2016 Sep 02 & 600 & Kbb & 100 \\ 
white light$^a$ & 2015 Sep 04 & 8$^b$ & Kbb & 50\\
\enddata
\tablenotetext{a}{Single column of lenslets illuminated}
\tablenotetext{b}{Note that the exposure time for individual white light scans varies based on mode; the exposure time given is used for the given mode only.}
\label{tab:obs}
\end{deluxetable}

\subsection{Example Science Data Sets}\label{sec:scidata}
Bright quasars and their host galaxies are a particularly useful data set for testing the DRP as they contain an extremely compact ($<$2 spaxels), bright point-like quasar with both continuum and strong emission lines, and underlying faint galaxy emission. A clean spectral extraction and high quality AO performance is needed for these data sets, since the science objective is to resolve the host galaxies of high-redshift (z$>$1) quasars, which typically span $<$1'' \citep{vayner2016providing}. 

On 2014 May 19 and 2015 August 9 OSIRIS observations were taken of the quasars 3C 298 (z=1.439; R=16.0 mag; H=14.5 mag) and 3C 9 (z=2.012; R=17.4 mag; H=15.6 mag) using the 100mas plate scale with the Hn3 ($\lambda_{cen}$=1.64 $\micron$) and Kn1 ($\lambda_{cen}$=2.01 $\micron$) filters, respectively. The LGS AO system was used for both. For 3C 298 there were a total of four dithered 600 second frames, while for 3C 9, there were a total of seven dithered 600 second frames. Separate sky frames were acquired during both observations. 

Initial data reduction in 2014 indicated potential issues with the DRP. The combination of the bright continuum source with strong emission lines made identification of spatial and spectral flux assignment artifacts easy. We further describe the QSO DRP results in Section \ref{qsodata}.

\subsection{Skies}
NIR sky spectra are useful for testing DRP effects on emission line spectra due to the abundance of narrow OH sky lines.  In addition to the OH emission lines, the sky spectrum exhibits a thermal background continuum, increasing in the K-band. Sky observations are frequently taken during the normal course of observing; however, several dedicated "deep skies" were taken in order to maximize the signal-to-noise. 

\subsection{Arc lamps}
Calibration frames using Ar, Kr, and Xe arc lamps were obtained. Emission lines were identified using a NIR line list\footnote{http://www2.keck.hawaii.edu/inst/nirspec/lines.html}. Arc lamp spectra are useful as a calibration source due to their high S/N emission lines, narrow line widths, and well-constrained wavelengths. 

Arc lamp spectra were obtained in two ways: first, with the entire lenslet array illuminated, as for most science exposures. In addition, arc lamp spectra were observed using a single lenslet column mask.  This mask consists of a slit wide enough to illuminate only one column of lenslets, which corresponds to one illuminated column of spaxels in the reduced data cube. On the detector, this produces spectra that are well-separated (i.e. separated by many times the FWHM of each spectrum) and thus flux on the detector can be assigned unambiguously to a single spaxel. The resulting data cube is expected to contain a single column of spaxels with full arc spectra, while the unilluminated lenslets correspond to dark spaxels with no spectra.

\subsection{Rectification Matrices}
\label{ssec:recmat}

\begin{figure}
\begin{center}
\includegraphics[width=\columnwidth]{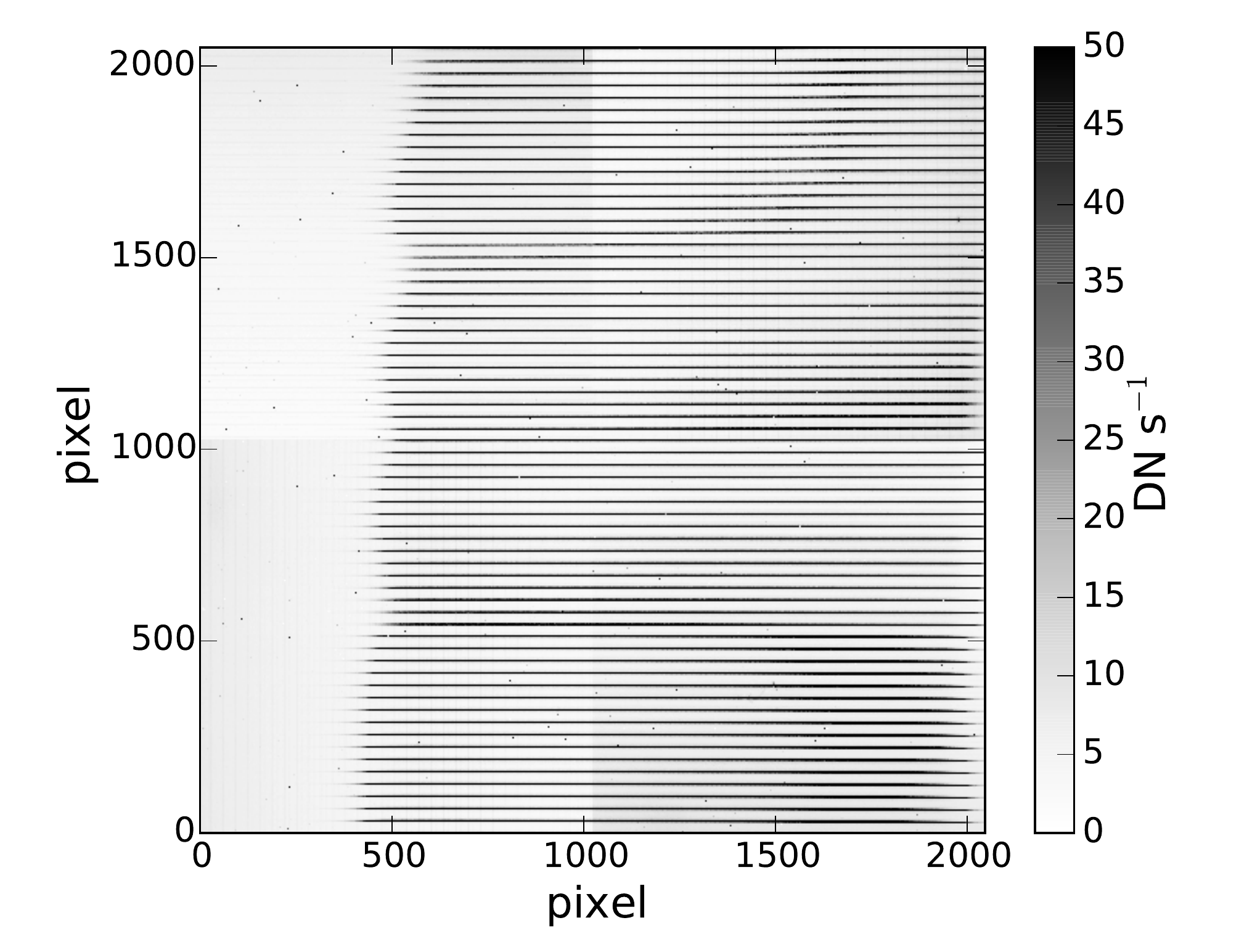}
\caption{Example single white light scan, used in the creation of a rectification matrix along with the rest of the set of white light scans. In this single scan, only one column of lenslets is illuminated, which produces well-separated traces. The x direction corresponds to the dispersion direction.}
\label{fig:whitelight}
\end{center}
\end{figure}

Flux on the detector is assigned to a spaxel in the data cube using an empirically determined matrix, the rectification matrix, of lenslet response curves. White light exposures are taken through the single lenslet column mask. The mask is stepped across the lenslet array, illuminating one column of lenslets at a time. The resulting frames display well-separated white light bands with no emission lines (see figure~\ref{fig:whitelight} for a single example scan). 

As the illumination source is uniform, the resulting flux on the detector is a function of lenslet response and pixel response only. These response curves are measured and encoded in the resulting rectification matrix. 

OSIRIS allows the selection of one of four spatial
samplings, from 0\farcs02 to 0\farcs1 per spaxel, in combination with one of 23 available filters, though not all filters are designed to work with all modes. This results in up to 88 modes that each require a separate set of white light exposures and a rectification matrix.

The rectification matrix file contains three extensions. The first extension is the y position (lower and upper bounds) of each spectrum on the detector, numbered in descending order from the top of the detector. The second extension is a quality flag. The spectra from some edge spaxels do not fall completely on the detector and thus are excluded from the final data cube; these are marked with a 0 in this extension. The third extension is the three-dimensional matrix containing the lenslet response curves. The first dimension is the width of the detector along the dispersion direction (2048 pixels), the second is the extent of the vertical response of each lenslet on the detector (16 pixels), and the third is the spectrum number (dependent on the mode selected). The lenslet response curves for the spectra flagged in the second extension are set to zero here. 

\section{FLUX ASSIGNMENT ARTIFACTS}
\label{sec:fluxassign}

The \textit{Spectral Extraction} DRP module (\texttt{spatrectif}) works to model the spectral PSF from each lenslet (which has a one-to-one correspondence with a single spaxel in the reduced data cube) using the average 1D PSF in the spatial direction, as further described in Section \ref{ssec:drp}. The current spectral extraction routine does not model the 2D structure of the PSF, which means that any asymmetries in the PSF will not be accounted for in the wavelength direction. Problems with the flux assignment algorithm were first noted beginning in December 2012, after OSIRIS moved to Keck-I and the new grating was installed. These issues manifested in reduced data cubes in both the \textit{spatial} direction, with artifacts affecting the total flux in neighboring spaxels, and in the \textit{spectral} direction, with artifacts appearing over a limited wavelength range in neighboring spaxels. In this section we describe the symptoms and problems seen with flux assignment artifacts with on-sky data sets and engineering data, and then follow with discussion on potential causes of the problem and future solutions.

\begin{figure*}
\begin{center}
\includegraphics[width=7.7in]{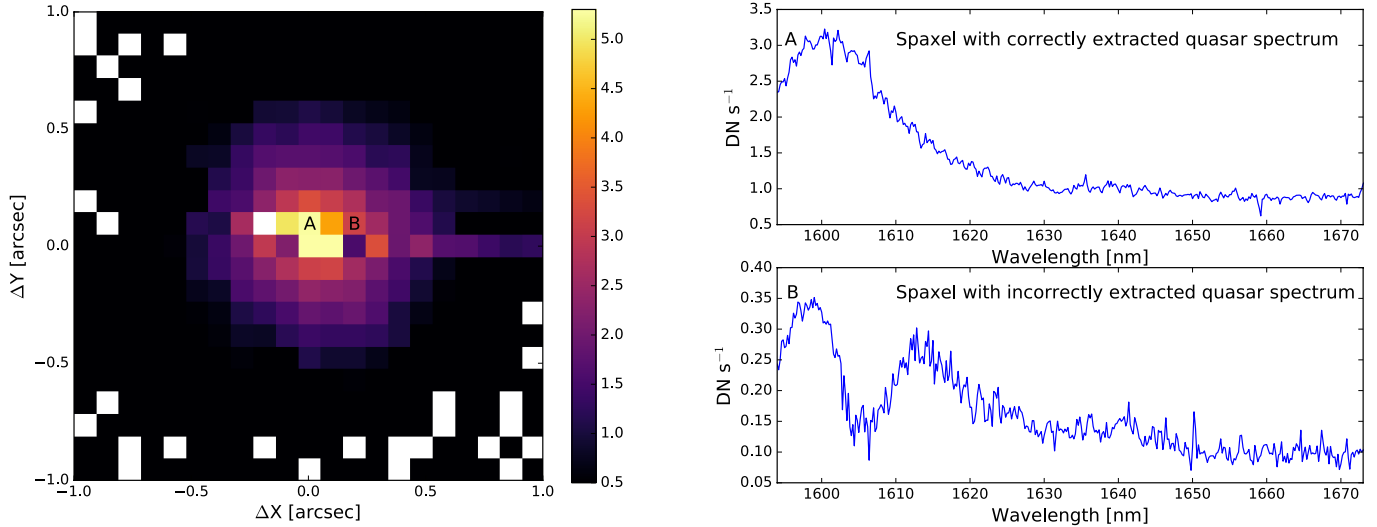}
\caption{Example of spatial flux assignment artifacts for quasar 3C 298. \emph{Left:} The data cube is collapsed in the wavelength direction to show spatial structure. The spatial distribution of quasar flux should show a smooth, nearly Gaussian PSF. Example spaxels A and B are marked. \emph{Top right:} Spaxel A shows the expected spectrum with characteristic broad-line H$\alpha$ emission. \emph{Bottom right:} The spectrum in spaxel B shows a large unphysical dip at 1605 nm, at the wavelength of the expected broad-line H$\alpha$ emission. This spaxel should be virtually identical to spaxel A and differ only by total integrated intensity. (Note that other spaxels in the central region, such as those with unusually low flux, also exhibit unusual spectra; spaxel B is chosen as a representative example.)}
\label{fig:OSIRIS_QSO_flux}
\end{center}
\end{figure*}

\subsection{Spatial flux assignment artifacts from pupil misalignment}\label{qsodata}
Spatial flux assignment artifacts are seen in QSO data taken in 2014 and 2015. The cause of these spatial artifacts is differences between the PSF of on-sky data and of the white light scans used to create the rectification matrices. This effect is apparent almost exclusively at the 100 mas plate scale. These differences are ultimately caused by misalignment between the on-sky pupil and that of the calibration unit, which feeds the Keck AO system and instruments with arc lamp and white light sources and which includes a simulated Keck pupil. Deviations between this calibration unit pupil and the on-sky pupil directly affect the shape of the PSF and the location of spectra on the detector. Data taken using the 100 mas plate scale show a small shift or translation between white light calibration spectra on the detector and that of on-sky science spectra. In addition, the PSF width of the white light calibration data is different from that of on-sky science data. We detail the investigation into the spatial flux assignment artifacts below. The on-sky data and the white light misalignment is still an ongoing issue and is not only limited to 2014 and 2015. We further discuss issues with the pupil in the 100 mas scale which may be the main cause of this misalignment.

The spatial flux assignment artifacts are evident in QSO data taken between 2014 and 2015 using the 100 mas plate scale (\S~\ref{sec:scidata}). The quasar, unresolved at this resolution, appears as a point source convolved by the spatial PSF in data cubes collapsed in the wavelength direction. The quasar spectra should be dominated by broad-line H$\alpha$ emission, represented by a broad Gaussian profile ($\sigma \sim$3000 km s$^{-1}$). Instead, in some individual spaxels, the broad-line H$\alpha$ emission has large discrepant dips. Other spaxels show other features that do not arise from the intrinsic science spectra. See figure \ref{fig:OSIRIS_QSO_flux} for an example using the 3C 298 reduced data cube (Hn3/100 mas). This 2014 data set signaled that there were major issues with the spectral extraction routine and flux assignment. 

\begin{figure*}
\begin{center}
\includegraphics[width=7.5in]{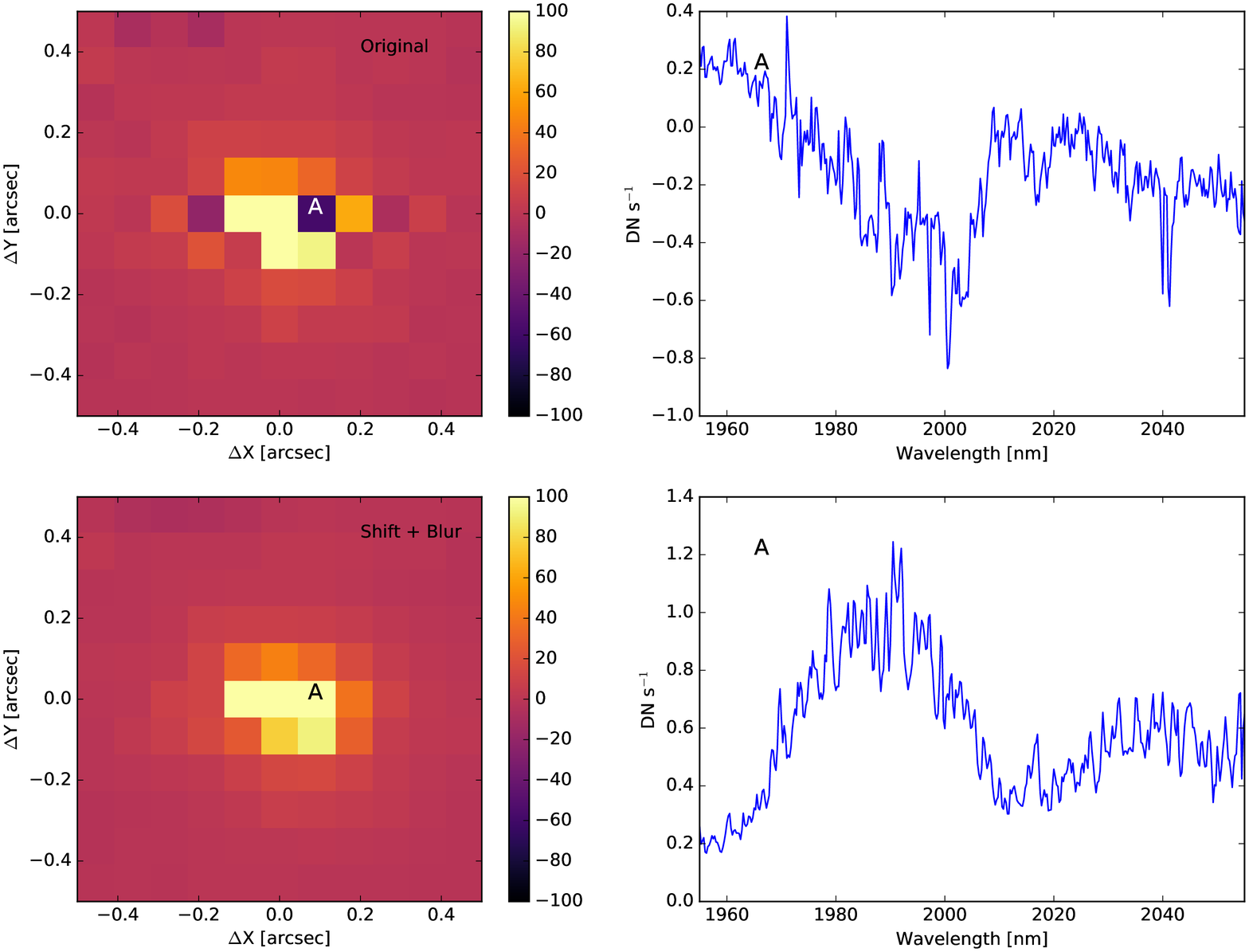}
\caption{Shifting and convolving the 3c 9 quasar data before reducing corrects the spatial flux assignment artifacts. \emph{Top panels:} The original data cube, collapsed in the wavelength direction (\emph{left}) shows discrepant low-flux spaxels, such as spaxel A, while individual spectra (\emph{right}) show unphysical features. \emph{Bottom panels:} Quasar 3c 9 data, after shifting and convolving the raw data to match the white light calibration data and rereducing. The \emph{bottom left} panel shows the data cube collapsed in the wavelength direction. The expected smooth spatial profile is now seen. The \emph{bottom right} panel shows the spectrum from spaxel A after applying the corrections, with the expected broad-line H$\alpha$ spectrum.}
\label{fig:OSIRIS_QSO_fix}
\end{center}
\end{figure*}

We investigate this effect by comparing the quasar continuum emission location on the 2D frame to that of the white light scan for the same lenslet. A Gaussian profile is fit to both the quasar continuum spectrum on the 2D frame and to the matched white light scan in the spatial direction, perpendicular to the dispersion direction. In the 2015 data set for quasar 3C 9 (Kn1/100 mas), the peak of the quasar continuum emission is translated by 0.53 pixels in the spatial direction from the white light scan at a given wavelength. In addition, the profile fitted to the white light scan is broader by $\sim$0.2 pixels compared to the on-sky profile. A shift is applied to the quasar data to align it with the position of the white light scans, and the quasar data are convolved with a Gaussian in the spatial direction to match the width of the white light scans. We show the results of this process in Figure \ref{fig:OSIRIS_QSO_fix}.

It is important to note that these spatial flux assignment artifacts seen in these tests with the QSO data are present almost exclusively in the 100 mas plate scale, where the on-sky pupil is slightly different than that of the calibration unit. OSIRIS's four spatial scales (0\farcs02, 0\farcs035, 0\farcs05, and 0\farcs1) are achieved by swapping in matched pairs of lenses that magnify the images onto the lenslet array, and have an effective pupil size matched to each beam size. This leads to differences in sensitivities and backgrounds for each of the four spatial scales and makes it less likely that spatial flux assignment artifacts will be observed in the other scales. 

Indeed, tests conducted comparing single lenslet column sky frames in both the 50 and 100 mas plate scales with matched white light scans show this discrepancy between on-sky data and calibration data. A Gaussian profile is fit at each illuminated spaxel in the reduced single lenslet column sky cubes and the matching reduced white light scan cubes, perpendicular to the illuminated lenslet column. The centroid and FWHM of the profile at each spaxel is compared between the sky and the white light scans. On average, the centroid positions differ by 0.03 spaxels in the 50 mas data and 0.14 spaxels in the 100 mas data. For the 50 mas data, the average FWHM difference between the sky and white light scans is 0.01 spaxels, while the average difference for the 100 mas data is 0.30 spaxels.

The cause of this mismatch is the differing pupil masks used for different plate scales. Each of the three fine scales (0\farcs02, 0\farcs035, 0\farcs05) have cold pupil stops mounted within the camera wheel, while the coarse scale (0\farcs1) has a fixed cold stop permanently mounted in the optical path\footnote{In 2006, the 0\farcs035 and 0\farcs05 scale pupil masks were redesigned, fabricated, and installed to match the mean aperture size of Keck (10.0 m) to lower the thermal background. These new pupil masks reduced the thermal background by $\sim$65\% with respect to the original pupil masks. In March 2008, we conducted a servicing mission for OSIRIS to install duplicate K broadband (Kbb) and narrowband (Kn3, Kn4, Kn5) filters in the filter wheel with new smaller pupil masks attached (9 m, effective). These K-band filters with their pupil masks are optimized to only work with the 0\farcs1 lenslet scale, and are referred to as Kcb, Kc3, Kc4, and Kc5 within the software, where ``c'' stands for ``coarse.''}. This has the unfortunate effect that the 0\farcs1 pupil must be oversized to ensure that when using other spatial scales their beams do not vignette. This was a known issue at the time of delivery, but will cause minor mismatch between the calibration and on-sky PSFs and higher thermal noise. Indeed the 100 mas reductions were cleaner and did not have these flux assignment artifact issues before 2012 when OSIRIS was on Keck-2 with the old grating. To date, we believe that the pupil change between Keck-2 and Keck-1 calibration unit may be a significant cause of the problem. 

\subsection{Spectral flux assignment artifacts from PSF asymmetry}\label{arclamps}
Spectral flux assignment artifacts are seen in bright emission line data, such as arc lamp calibrations or sky frames. These artifacts are caused by the discrepancy between the real 2D PSF of a source on the detector and the smoothed PSF assumed during the construction of the rectification matrix. The smoothed PSF effectively averages the real 2D PSF of every pixel into a one-dimensional line spread function (LSF) perpendicular to the spectral dispersion direction. For example, the 2D PSF of a bright source on the detector, such as an emission line, exhibits some asymmetries due to imperfections in the optical system. For the OSIRIS optical system, the real 2D PSF has a coma, or is flared, towards the left. However, the rectification matrix is constructed using white light scans to estimate the LSF perpendicular to the dispersion direction. It then uses this LSF estimate to assign flux iteratively from the detector into the correct spaxel, which effectively smooths or averages the PSF at each spectral channel. In effect, the white light LSF is narrower than the real PSF across the flared side and wider across the non-flared side. The width difference between the LSF and the 2D PSF introduces an error into the flux assignment algorithm and produces the flux assignment artifacts. We demonstrate this issue below.

\begin{figure}
\begin{center}
\includegraphics[width=\columnwidth]{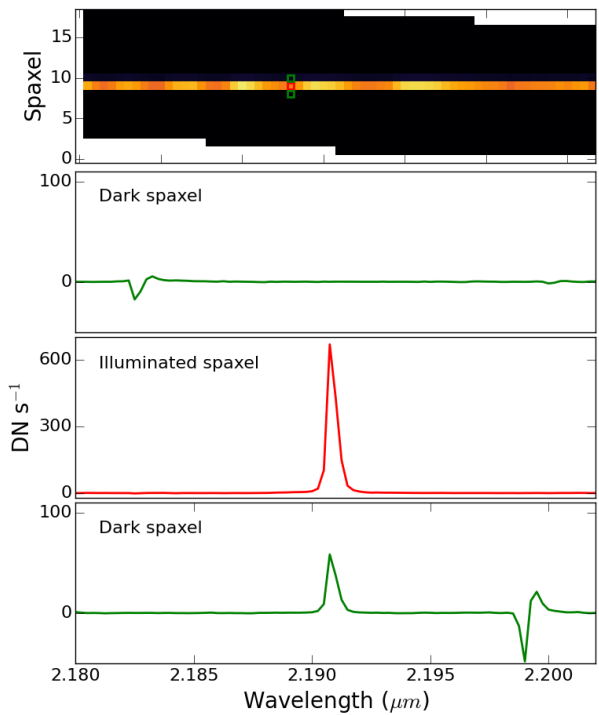}
\caption{Example of flux assignment artifacts in a September 2016 single-column arc lamp spectrum (Kbb/35 mas). Flux is conserved in the correct spaxel, but flux assignment artifacts are visible at $\pm$32 spectral channels in adjacent spaxels. \emph{Top panel:} Spectral channel map at 2.191 $\mu m$, the peak of a bright Kr emission line. Only one column of lenslets is illuminated. The green boxes mark the two dark spaxels whose spectra are shown at bottom, while the red box marks the shown illuminated spaxel. \emph{Bottom panels:} Segment of a spectrum in the three adjacent spaxels indicated in the channel map. \emph{Second panel:} Dark spaxel adjacent to an illuminated spaxel. This spaxel should show no flux, but negative artifacts are visible. The artifacts are offset by -32 spectral channels from the emission line in the middle panel. \emph{Third panel:} Illuminated spaxel. \emph{Bottom:} Dark spaxel adjacent to an illuminated spaxel. This spaxel should not show flux, but positive and negative artifacts are visible. The blue artifacts are at the same spectral channel as the emission line in the middle panel, while the red artifacts are offset by +32 spectral channels.}
\label{fig:misflux}
\end{center}
\end{figure}

Issues with the flux assignment algorithm in the spectral direction are most evident in the presence of a bright emission line. The flux assignment artifacts manifest as increased systematic error in neighboring spaxels, shifted by 32 spectral channels away from the line. Figure~\ref{fig:misflux} shows an example using a single-column arc lamp spectrum. 

\begin{figure}
\begin{center}
\includegraphics[width=\columnwidth]{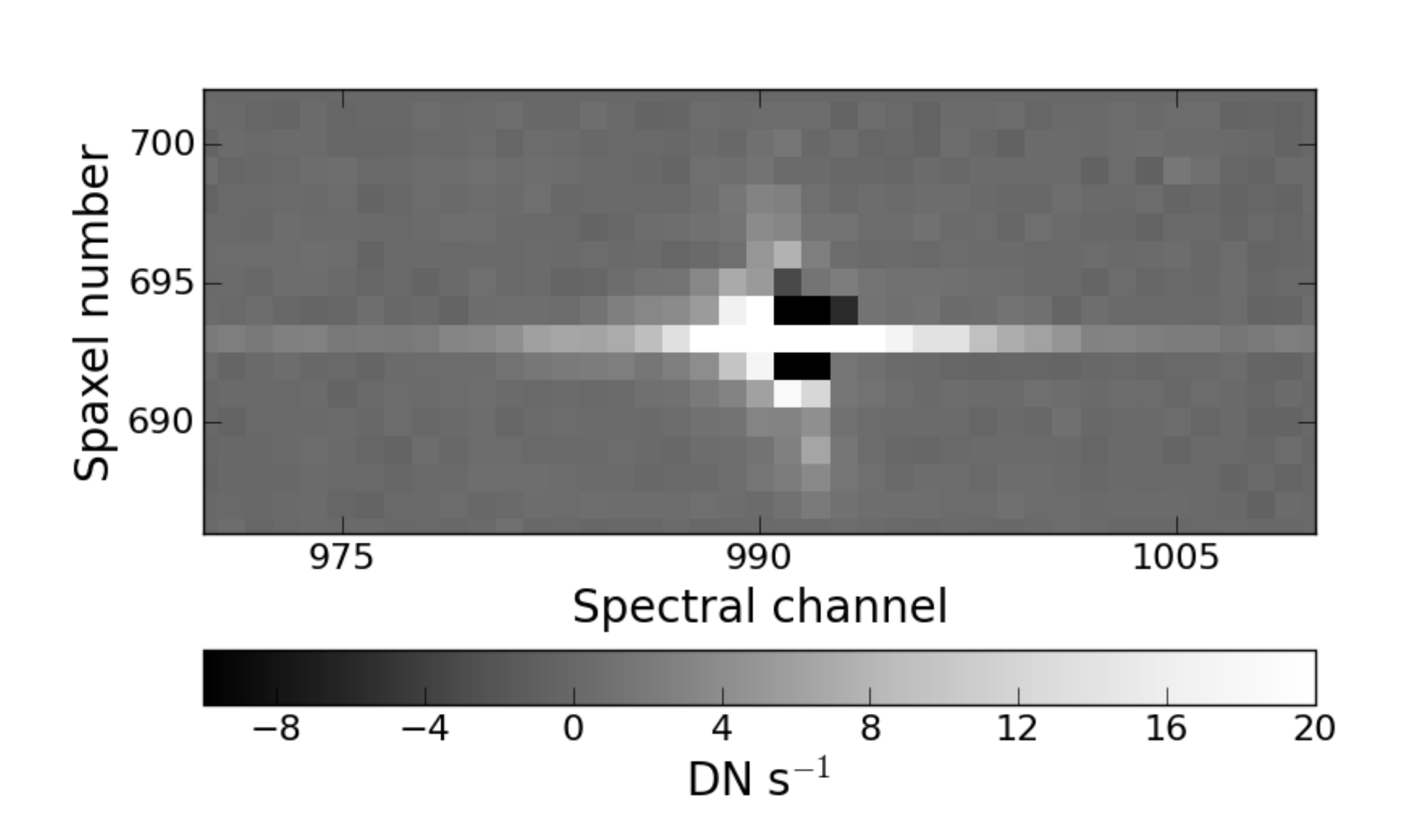}
\caption{Example of flux assignment artifacts, shown after the raw flux has been assigned to a single spaxel but before wavelength calibration or assembly into a cube. Note that as the wavelength calibration is not yet applied, the wavelength direction is opposite from the usual sense, and increases towards the left. A section of a September 2016 reduced Kbb/50 mas single-column arc lamp frame centered on a bright emission line is shown. Each row corresponds to an individual spaxel, while each column corresponds to a spectral channel. The wavelength solution has not yet been applied so there is not a one-to-one correspondence between channel number and wavelength. Neighboring spaxels are shifted by 32 spectral channels with respect to each other. In this example, all flux should fall in spaxel 693. Residual flux above and below the bright emission line at roughly channel 990 is evidence of flux assignment artifacts. The shape of the artifacts reflects the shape of the 2D PSF.}
\label{fig:swapchan}
\end{center}
\end{figure}

The 32 channel shift is due to the stagger in wavelength between neighboring spaxels; when the spectra from the 2D frame are assembled into the data cube and the wavelength calibration is applied, this translates into a shift of 32 spectral channels in neighboring spaxels. The flux assignment artifacts are thus more easily pictured without this wavelength calibration applied. Figure~\ref{fig:swapchan} shows an example using a segment of a single-column arc, centered on a bright emission line. Positive and negative artifacts are evident both above and below the emission line in the illuminated spaxel. These are the cause of the artifacts in the dark spaxels seen in figure~\ref{fig:misflux}.

\begin{figure}
\begin{center}
\includegraphics[width=\columnwidth]{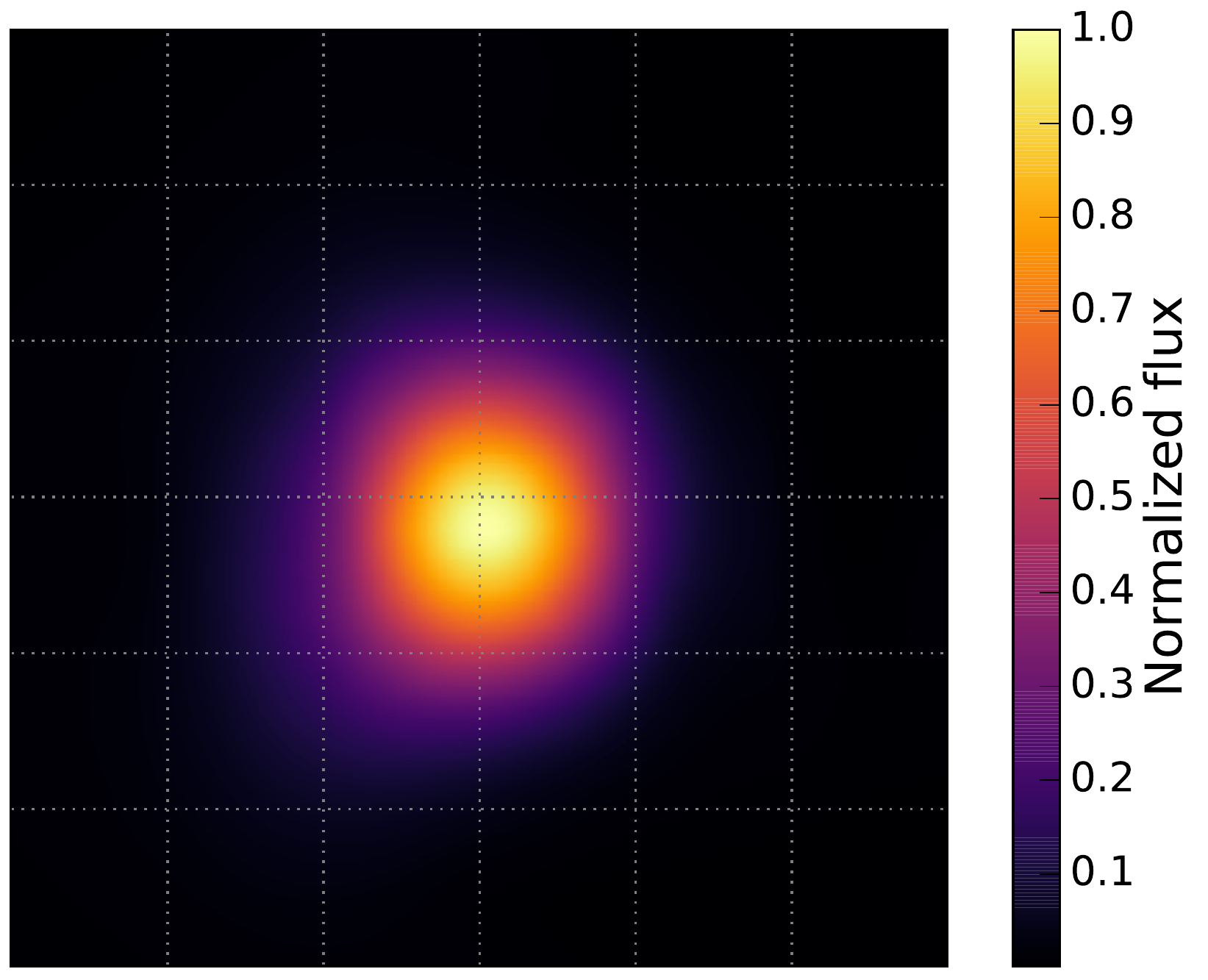}
\caption{PSF on the detector, supersampled by a factor of 100. Gray lines represent the detector pixel boundaries, for scale.}
\label{fig:PSF}
\end{center}
\end{figure}

The positive and negative artifacts around the bright emission line show a clear asymmetry---positive artifacts are preferentially seen redward of the emission line peak while negative artifacts appear blueward of the line peak. This artifact shape corresponds with the PSF of lines on the 2D frame. Figure~\ref{fig:PSF} shows the supersampled PSF of a bright arc lamp emission line, which was constructed by stacking the raw images of this line in all spaxels in a single 2D frame and supersampling it by a factor of 100. Note that as there is some variation in the 2D PSF FWHM as a function of lenslet and wavelength \citep{boehle2016upgrade}, a more robust determination of the PSF shape would come from stacking multiple matched images of a given emission line in the same lenslet to remove these effects. However, given the limitations of available calibration data, this analysis is left for future work. The PSF, shown in figure~\ref{fig:PSF} and effectively averaged over the detector, is asymmetric and flared towards the left, likely reflecting a coma from the instrument optics. The correspondence between the flux assignment artifacts and the average PSF reflects that the cause of the artifacts is the asymmetry of the PSF. 

From testing, it appears that flux is conserved in individual spaxels in the presence of spectral flux assignment artifacts. Three tests were conducted using single-lenslet column arc lamp data from 2015 and 2016. In all tests, a circular aperture is placed on a bright emission line in the 2D detector frame, before flux assignment. The spatial rectification module is run, but the DRP is stopped before the wavelength solution is applied, creating output similar to figure~\ref{fig:swapchan}. A rectangular aperture is placed on just the illuminated lenslet (e.g. spaxel 693 in the example in figure~\ref{fig:swapchan}) to remove the effect of the artifacts. The enclosed flux is compared before and after the flux assignment. In all tests, the enclosed flux after the flux assignment matches within 3$\sigma$, and in two of the three tests, the flux matches at the 1$\sigma$ level. As these tests show that roughly the same amount of flux appears in the integrated emission line in the final data cube, the positive and negative artifacts must effectively cancel each other out. 

However, this issue does induce extra error due to artifacts from both emission lines in the science data and from OH sky lines. In addition, the level of the artifacts, and the induced error, has changed with time. We create a metric to measure the level of the flux assignment artifacts with time. We measure the absolute value of the peak of the flux assignment artifacts resulting from a bright emission line as a percentage of the peak of the line itself. As the artifacts can be asymmetric, we measure the artifacts separately in the two neighboring spaxels. We calculate this metric separately for both arc lamp and sky spectra. As the sky spectra are generally much lower in S/N than the arc lamps, we calculate the metric on a median spectrum of a column of spaxels in the sky cube. For the arc lamps, the metric is calculated separately on each individual spaxel. We report the mean values and their standard deviations from our investigation for each observation type and time period in table~\ref{tab:flux}.

\begin{deluxetable*}{lccccc}
\renewcommand{\arraystretch}{0.9}
\tablecaption{Flux Assignment artifacts}
\tablewidth{0pt}
\tablehead{
\colhead{} & \colhead{} & \colhead{} & \colhead{} & \twocolhead{Absolute value of peak artifacts}  \\
\colhead{Period} & \colhead{Data type} & \colhead{Modes} & \colhead{No. of frames} & \colhead{-1 spaxel} & \colhead{+1 spaxel}
}
\startdata
2012--2015 & arc & Hbb/20; Kbb/35, 50, 100 & 6 & 7.2$\pm$3.7\% & 5.3$\pm$1.8\% \\
2012--2015 & sky & Hbb/50; Kn3/20, 35, 50, 100 & 6 & 2.4$\pm$0.9\% & 3.0$\pm$1.0\% \\
2016--present & arc & Kbb/20, 35, 50, 100 & 4 & 2.2$\pm$0.6\% & 5.9$\pm$0.5\% \\
2016--present & sky & Jn2/35; Jbb/50; Hbb/50; Kbb/35, 50, 100 & 6 & 4.8$\pm$5.1\% & 6.1$\pm$2.1\% \\
\enddata
\label{tab:flux}
\tablenotetext{}{}
\end{deluxetable*}

Prior to December 2012, the artifacts were much smaller but this issue became more prominent after the grating upgrade. Results from the analysis detailed in this section were taken into account during the most recent hardware upgrade, the detector replacement in 2016. To mitigate both the spectral flux assignment artifacts and the spatial rippling (see next section, \S~\ref{sec:spatrip}), the PSF elongation was rotated 90 degrees to be slightly broader in the dispersion direction. This rotation eased the discrepancy between on-sky data and the white light scans, and has qualitatively lessened the degree of the spectral flux assignment artifacts. This is also demonstrated by the reduction in the level of the arc lamp artifacts after the detector replacement (Table~\ref{tab:flux}). (The sky line measurements are much noisier, but are consistent with no reduction in the level of the flux artifacts.)  However, it is still a stronger effect than it was prior to 2012.

For observers concerned about the spectral flux assignment artifacts, flux appears to be conserved, but additional systematic error at the level of a few percent of the peak emission line flux (Table \ref{tab:flux}) may be a concern. In addition, the placement of the artifacts, if they fall on a spectral feature of interest, may impact science measurements. From our investigations, the ultimate cause is likely the method of assigning flux to spaxels (e.g. the white light scans and rectification matrices). \citet{boehle2016upgrade} show that the FWHM of individual bright emission lines in the 2D detector image vary by up to a few tenths of a pixel in both the X and the Y direction as a function of lenslet and wavelength. Thus to fully resolve this issue, an implementation of a two-dimensional PSF estimation and spectral extraction algorithm may be necessary, though the task is non-trivial. We would need a 2D estimate of the PSF on the detector for every spaxel and wavelength combination. However, as the internal structure of OSIRIS is fixed, we can currently only sample the 1D LSF. Work is underway to investigate whether existing measurements of the 2D PSF variation in OSIRIS are adequate to improve improve the spectral extraction. Alternatively, the CHARIS IFS on Subaru utilizes a tunable laser to sample the 2D PSF at all wavelengths \citep{brandt2017data}; something similar may need to be installed on OSIRIS to implement a 2D flux extraction algorithm.

\section{SPATIAL RIPPLING}
\label{sec:spatrip}

\begin{figure*}
\begin{center}
\includegraphics[width=\textwidth]{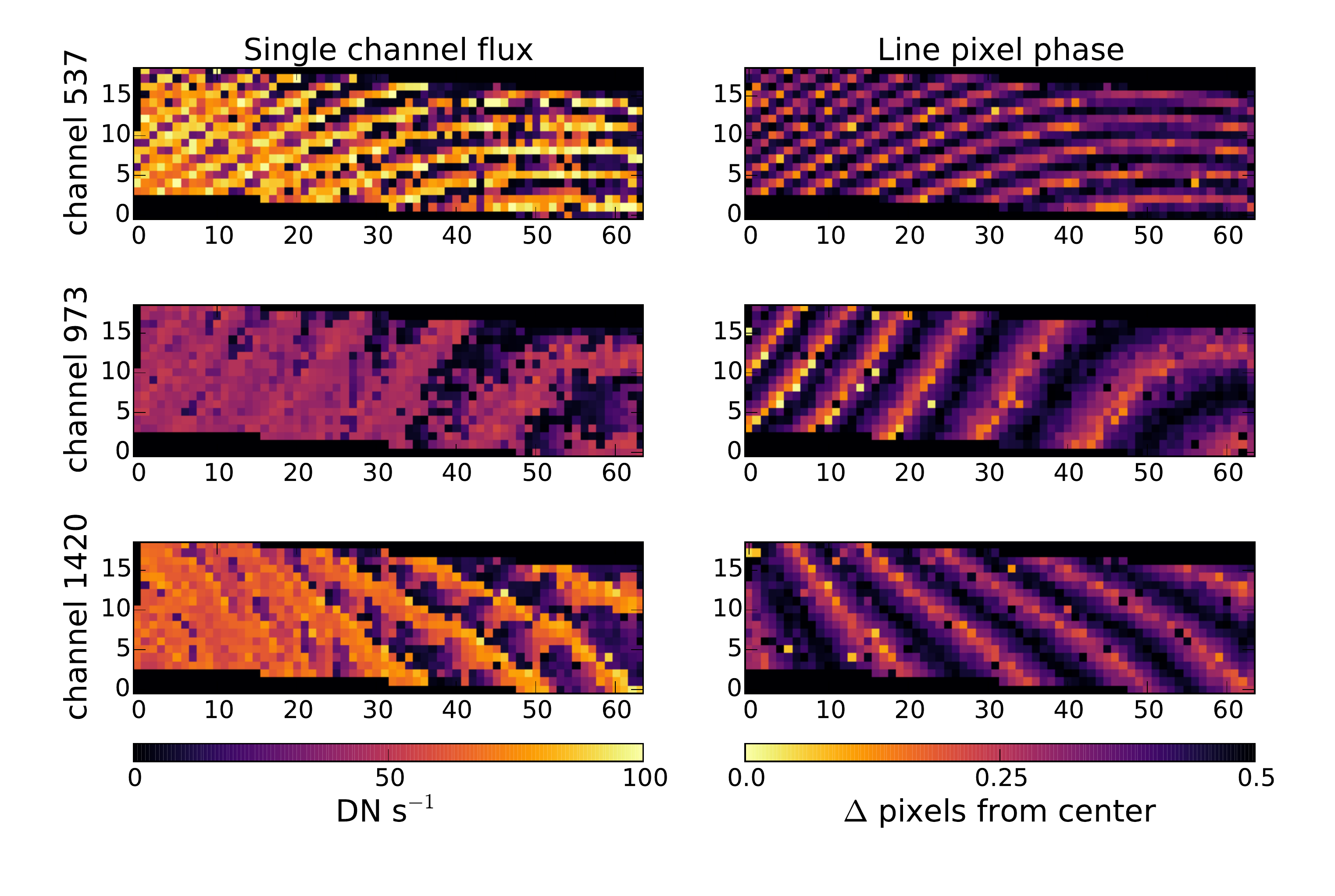}
\caption{A pattern of spatial rippling, best seen in bright emission line single-channel maps, is caused by the undersampled PSF in the dispersion direction on the 2D frame. \emph{Left:} Single channel maps from a reduced Kbb/50 mas arc lamp cube from December 2015. Selected channels are chosen at the peak of several bright emission lines. \emph{Right:} Pixel phase, or the position of the peak of the PSF in a given pixel, of the emission lines (\emph{left}) along the dispersion direction on the detector. Units are the absolute value of the pixel phase; lighter colors indicate a more centered emission line, while darker colors indicate an emission line whose peak falls near the edge of a pixel.}
\label{fig:ripphase}
\end{center}
\end{figure*}

Individual spectral channel maps in the reduced OSIRIS cubes show spatial rippling, or a pattern of brighter and fainter spaxels unrelated to incident flux. The spatial rippling is especially apparent for bright emission lines, is wavelength dependent, and changes rapidly with wavelength. Figure~\ref{fig:ripphase}, left panels, shows an example of this rippling. Each panel is a 2D cut at a single wavelength channel in a reduced arc lamp cube. The three channels shown are each centered on a bright emission line in this bandpass.

The rippling pattern is roughly consistent at a single spectral channel across filters and pixel scales, with some magnification, and stays roughly consistent with time. However, the scale of the rippling, or the magnitude difference between the bright and faint spaxels in the pattern, has changed with time. In particular, it was much less evident before the grating upgrade in 2012.

The spatial rippling is 
caused by the subsampled PSF on the detector. Figure~\ref{fig:ripphase}, right panels, shows the pixel phase of the bright emission lines in the corresponding left panels. These pixel phases, or the position or centeredness of the emission line PSF on a single pixel, are measured in the dispersion direction on the 2D frame, with lighter colors representing more centered emission lines. 

The correspondence between the pixel phase of a given emission line and its channel map is due to the narrowness of the PSF on the detector. After the 2012 grating upgrade, the focus on the detector was sharpened along the dispersion direction, leading to a PSF that is subsampled in the dispersion direction. Emission lines that are centered on a pixel have more of their flux assigned to a single spectral channel in the reduced cube, while lines that fall at the edge of a pixel have their light split across two channels. This leads to the rippling in the reduced cube channel maps. Before the grating upgrade, the PSF was broader and was fully sampled by the given pixel scale, so this rippling effect was not present.

\begin{figure}
\begin{center}
\includegraphics[width=\columnwidth]{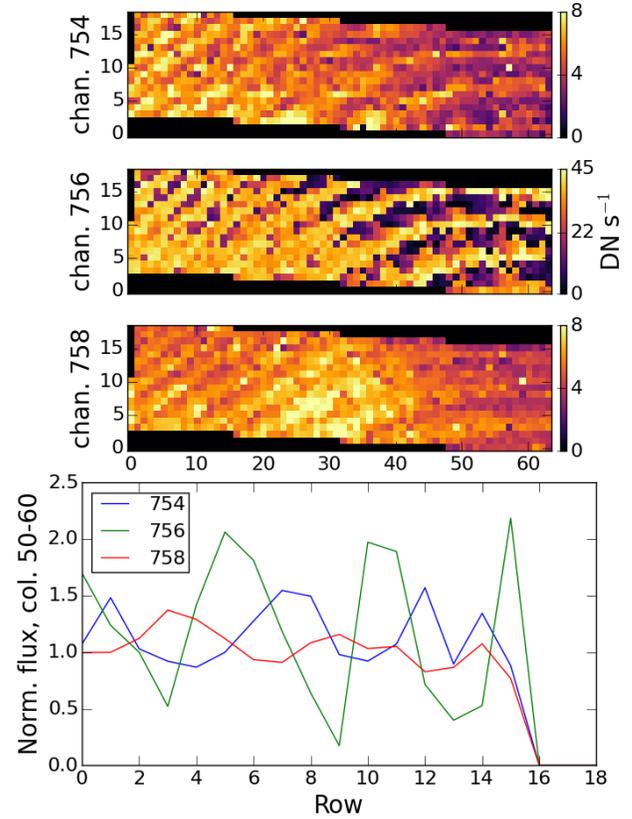}
\caption{The spatial rippling patterns change from channel to channel. \emph{Top:} Channel maps across a single emission line, at the peak of the line and at 2 wavelength channels away on either side. The change in rippling pattern with wavelength is quite apparent, particularly when comparing the right sections of the top and bottom panels. \emph{Bottom:} Vertical cuts across all three channel maps, taken between columns 50 and 60 (columns have been median combined). The cuts have been normalized for display. The vertical position of the horizontal stripes in each channel are shifted by 1--2 rows with respect to each other.}
\label{fig:spatripline}
\end{center}
\end{figure}

The rippling pattern can affect science data, particularly because the pattern changes rapidly. Figure~\ref{fig:spatripline} shows three single channel maps, close in wavelength, across one arc lamp emission line. Though the maps are separated by at most 5 wavelength channels, the shift in the rippling pattern is visible by eye. This change of pattern with wavelength induces extra noise into the cube and may affect measurements such as kinematics or channel maps. 

\begin{figure*}
\begin{center}
\includegraphics[width=\textwidth]{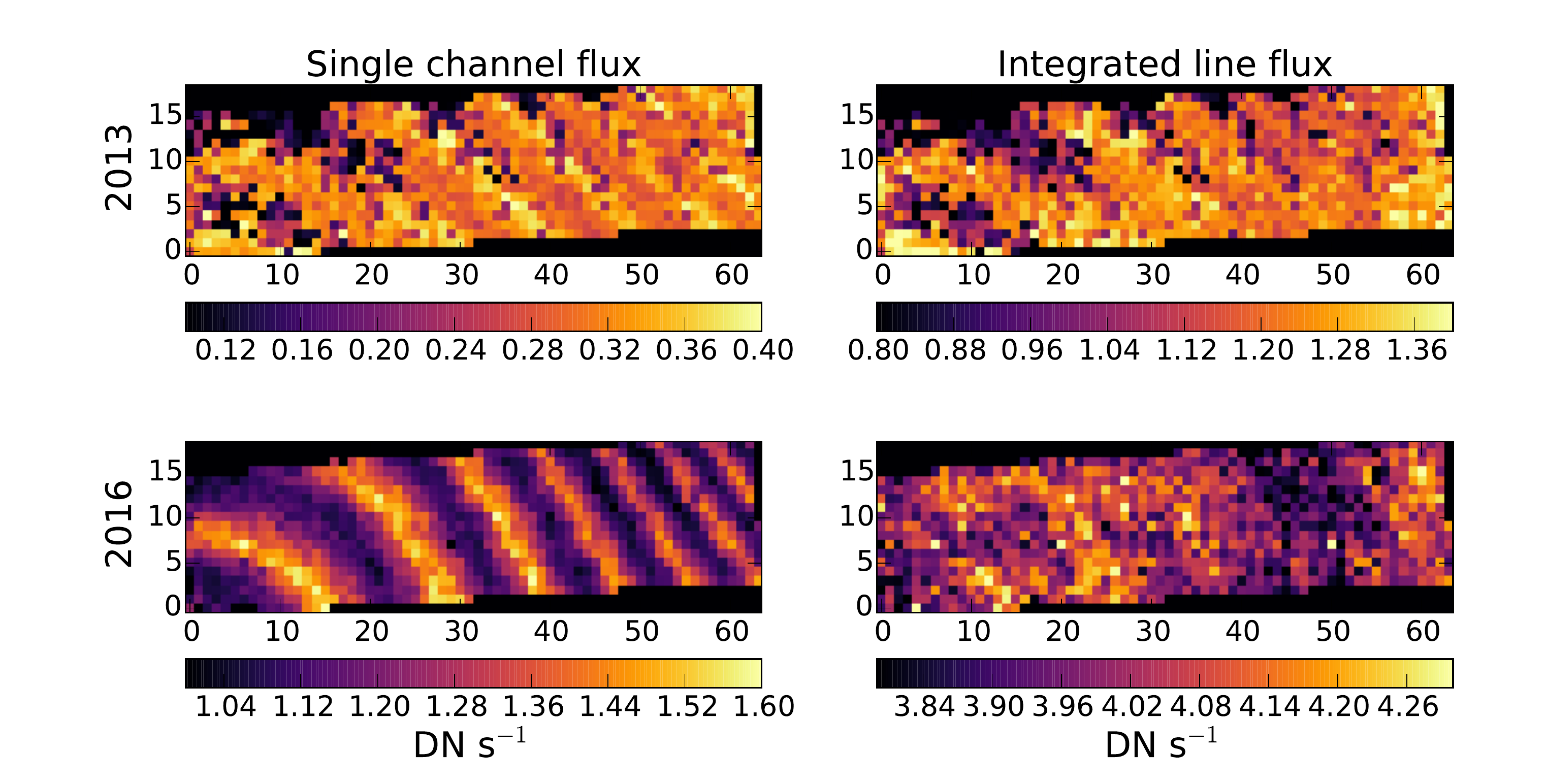}
\caption{\emph{Left:} Single channel maps of a OH line (2.196 $\mu m$, channel 925) from a reduced Kbb/35 mas sky cube from May 2013 (\emph{top}) and May 2016 (\emph{bottom}). \emph{Right:} integrated flux over 10 channels (920--930). }
\label{fig:skyripp}
\end{center}
\end{figure*}
 
While individual channel maps show a strong rippling pattern, integrated line flux is conserved in data from most epochs. However, data taken in the period from December 2012 to December 2015 is subject to a stronger spatial rippling pattern that does not resolve after integrating over the emission line. Figure~\ref{fig:skyripp} shows the peak and integrated line flux of an OH sky line in frames taken in 2013 and in 2016. In 2016 (bottom panels), while the rippling pattern is evident in single channel maps, the integrated line maps are smooth. On the other hand, integrated emission line maps from 2013 data (top panels) show spatial rippling in both the single channel and in the integrated flux maps. The cause of the persistent rippling seen in integrated flux maps from 2012 to 2015 is likely due to differences in the PSF width as compared with previous or later periods. The PSF in the dispersion direction is slightly narrower in the period between December 2012 and December 2015 than in the current configuration; during the 2016 detector upgrade, the PSF elongation was rotated so as to be broadened slightly in the dispersion direction. Correspondingly, during the period from December 2012 to December 2015, the PSF perpendicular to the dispersion direction was wider. The confluence of the flux assignment algorithm issues plus the PSF subsampling likely caused the rippling seen in the integrated flux maps.

In most data epochs, the spatial rippling issue is a nuisance but does not generally affect science data or calibrations\footnote{Note that the wavelength solution is calculated using 2D data, as described in \S~\ref{sec:wavesol}, and is thus unaffected by this rippling effect.}. Integrated line maps do not contain a residual rippling pattern, though single channel maps may, if mapping a narrow emission line. If emission lines are intrinsically broadened, as for some science cases, the lines may be sampled sufficiently to produced smooth single channel maps without integration. However, caution must be used in two cases. First, if the sky emission is not subtracted using a separate sky frame, but is instead extracted from a section of the science data cube, the spatially varying rippling pattern will not subtract out and will induce artifacts at the wavelengths of the sky lines. Second, data taken between December 2012 and December 2015 containing narrow emission lines from resolved science targets will show this rippling pattern, even in integrated channel maps. Testing using the sky frames shown in figure~\ref{fig:skyripp} shows that the standard deviation of an integrated line map relative to the median of the frame is 15\% in 2013, versus 5\% for the 2016 integrated line map. 

\section{COSMIC RAYS IN THE RECTIFICATION MATRICES}
\label{sec:crinrecmat}
\begin{figure*}
\begin{center}
\includegraphics[width=\textwidth]{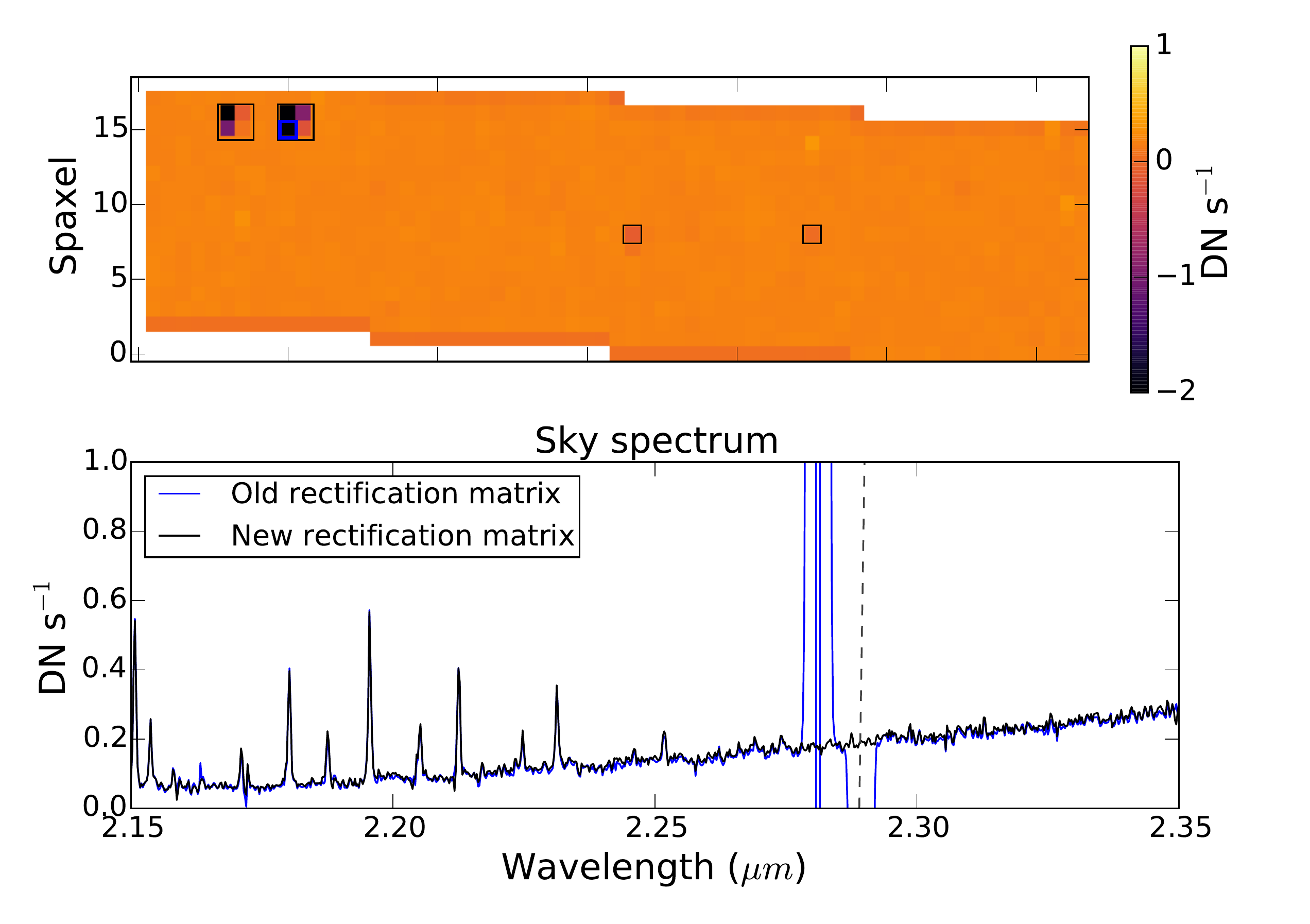}
\caption{Old rectification matrices did not properly account for cosmic rays; these led to artifacts in the reduced data cubes. These artifacts can be seen here in a reduced sky cube. \emph{Top:} Channel map (single wavelength slice) from a reduced sky cube. The spaxel shown in the bottom panel is boxed in blue. Other spaxels contaminated by improper cosmic ray removal in the rectification matrix are boxed in black. The color stretch has been adjusted to show the negative residuals due to the improperly removed cosmic rays. \emph{Bottom:} Section of the spectrum from the reduced sky cube, at the position shown by the blue box in the top panel. In blue is the spectrum from the cube reduced using the old rectification matrix, still contaminated by cosmic rays. In black is the same spectrum, reduced using the new rectification matrix that has been corrected for cosmic rays. The dashed vertical line shows the spectral channel represented by the map at top.}
\label{fig:crspikes}
\end{center}
\end{figure*}

The publicly available rectification matrices suffer from inadequate cosmic ray removal in the corresponding white light scans, which results in large flux spikes in the final reduced data cubes (figure~\ref{fig:crspikes}, bottom panel). These spikes occur in just a few spectral channels in a few spaxels, but display flux values up to 10$^8$ DN s$^{-1}$ (both positively and negatively signed), while the maximum real flux in a single spectral channel in one spaxel of a 900 s sky frame is on the order of 1 DN s$^{-1}$. In addition, the saturation limit of the detector is 33,000 DN (pre-2016) or 65,535 DN (2016--present), so such high counts are unphysical.

Prior to 2015, the rectification matrices were created using a single white light scan at each position of the lenslet column mask, disallowing cosmic ray removal by median frame combination\footnote{Single scans at each position were used because taking these scans is a very time-consuming process. A scan must be taken for every lenslet column for each of the 88 filter and plate scale modes. In order to achieve an adequate S/N, scans at the smaller plate scales and with some filters require exposure times of up to 30s.}. To remedy the inadequate cosmic ray removal, new rectification matrices were created in 2015 using a set of three white light scans at each lenslet column position. The scans at each lenslet column position are medianed together to robustly remove cosmic rays (Randy Campbell, private communication). The matrix made with the medianed scans produces a data cube that is free from spikes unrelated to instrumental variation. Rectification matrices using the median method were created for each OSIRIS mode in 2015, covering the period between December 2012 and December 2015, and are publicly available on the OSIRIS instrument website\footnote{http://tkserver.keck.hawaii.edu/osiris/}. All new rectification matrices, including for the period beginning in January 2016, are created using this median scan method.

New white light scans and new rectification matrices are created each time the instrument is modified and are unique to its physical setup. Therefore, data taken prior to the physical configuration beginning in December 2012, when this issue was discovered, must be reduced using a rectification matrix that still suffers from cosmic ray effects. Cosmic rays are not easily removed from the original scans, as the real flux peaks can be sharply peaked and sigma-clipping routines remove too much real flux. However, the rectification matrix itself is simply a three-dimensional array and any cosmic rays are easily identified and interpolated over. 

\section{BAD PIXEL MASK}
\label{sec:badpix}
We have determined that there are two types of bad pixels that lead to artifacts in the reduced data cubes: bad pixels on the detector (dead and hot pixels) and cosmic ray hits. The number of bad pixels on the detector far outnumber the typical number of cosmic rays in a 600 to 900 s exposure.\footnote{Hot pixels are expected to be less of a concern for the white light scans used for the rectification matrices. The white light scan exposures are short; the exposure time varies by mode, but is generally on order of a few seconds. The white light lamps are also bright, and thus the hot pixels are expected to be much fainter than the incident white light lamp flux. The cosmic rays are removed from the white light exposures as outlined in \S~\ref{sec:crinrecmat}.} Generally about 98 to 99\% of bad pixels are static and about 1 to 2\% are from cosmic rays. We developed a method for identifying hot or dead pixels on the detector and have included a script in the pipeline to apply the bad pixel mask to raw data. We discuss below our method for identifying hot and dead pixels.

Hot pixels are pixels on the detector that have higher than normal counts compared to a typical pixel given the same incoming flux. These pixels can lead to large artifacts in the reduced data cube, reducing the overall signal to noise. These are a source of systematic error in spectral features. 

\begin{figure}
\begin{center}
\includegraphics[width=\columnwidth]{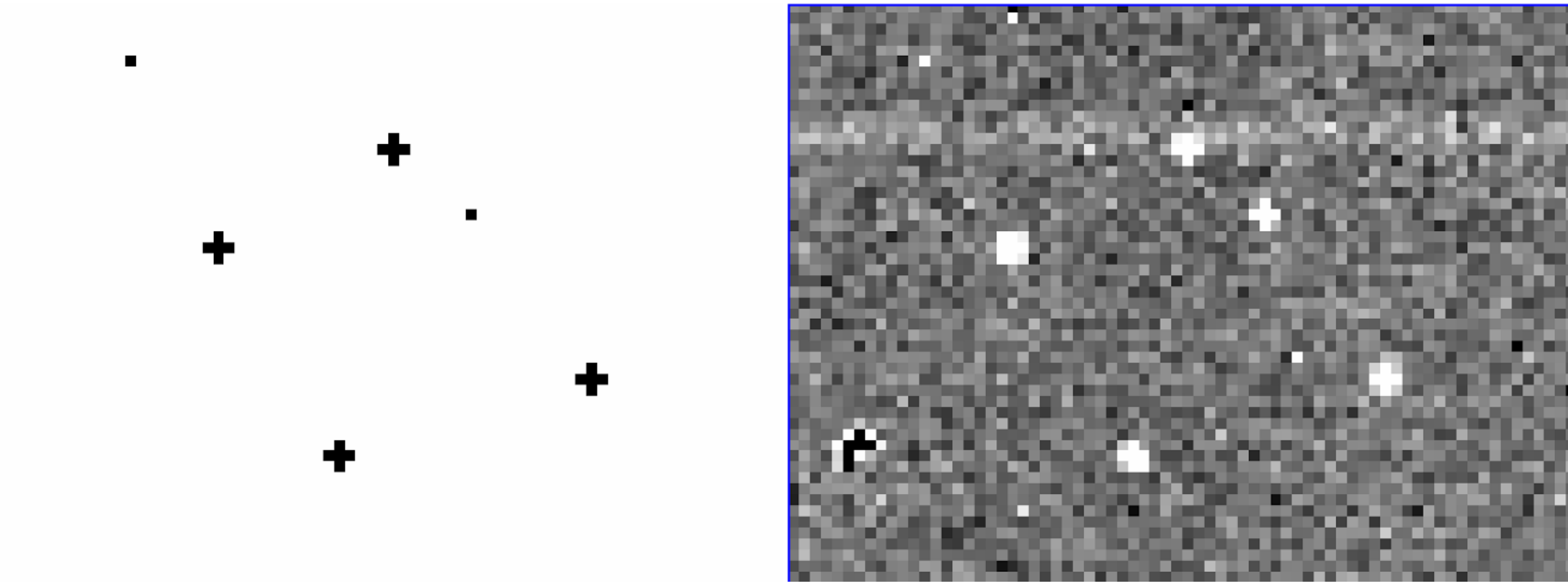}
\caption{Example of the hot pixel mask (left) generated from the median dark frame (right). These hot pixels are significantly higher compared to the median dark pixel value. Note that the discolored pixels in the bottom left of the median dark frame are dead pixels and are masked out in a separate step.}
\label{fig:hotpix}
\end{center}
\end{figure}

To detect hot pixels, we created a median of 5 darks with 900 s integration time from 2017 Sep 2. Using this median dark, we identified pixels that significantly deviate from the median value of pixels in the dark frame. We developed two hot pixel masks, one with a threshold of 50$\sigma$ above the median value and one with a threshold of 15$\sigma$ above the median value. At the 50$\sigma$, about 4000 hot pixels are detected (or approximately 0.1\% of total detector pixels), while at 15$\sigma$ about 10,000 hot pixels are detected (0.24\%). The median dark frame and a sample hot pixel mask are shown in figure~\ref{fig:hotpix}. 

We tested this mask by applying it to raw data. After reducing the data cube, we compared it to the cube created without the mask. These tests show that the mask does indeed reduce the number of large single-pixel spikes in the final data cubes. The tests also show that the bad pixel mask introduces some artifacts. A large majority of these artifacts are at a much lower level than the pixel values that exist when the bad pixel mask is not applied. In general, the artifacts that are removed are at least 10 times larger than the artifacts that are created. In addition, using the bad pixel mask removes twice as many artifacts as it creates. Overall, using the new bad pixel mask reduces noise in reduced test data by roughly 5\%. This bad pixel mask can now be optionally applied by users of the pipeline. 

\section{WAVELENGTH SOLUTION}
\label{sec:wavesol}
The wavelength solution is used to assemble the wavelength vector on a spaxel-by-spaxel basis; it converts a given x-position on the detector to a wavelength in the assembled data cube. The wavelength solution is determined empirically by calculating a third-order polynomial fit to the positions of arc lamp emission lines in each spaxel on the detector image. This fit is then used when interpolating the detector flux into wavelength space in the final reduced cube. In addition to calibration lamp measurements, corrections from instrumental effects such variations in the wavelength solution with temperature are also applied to the wavelength solution on each specific data cube. 

Scientifically, accurate wavelength solutions are important for high S/N radial velocity measurements. For example, measurements of the orbits of binaries or stars at the Galactic center requires an accurate wavelength solution to derive physical properties such as the stellar mass or the mass of the supermassive black hole \citep[e.g.][]{2016ApJ...830...17B}. Currently, the accuracy of the wavelength solution in the Kn3 filter, where the Br$\gamma$ line is located, is about 0.1\AA, or 2 km/s at Br$\gamma$. As a function of resolution, this is a roughly 3\% error. 

The advantage of a Nasmyth mounted instrument like OSIRIS is that calibrations such as the wavelength solution are generally very stable, even over a period of years, if the instrument is unopened. However, when the instrument is opened for servicing, the wavelength solution generally needs to be rederived. The typical shift in the wavelength solution after a cycle of warming, opening, servicing, and cooling the instrument is 2--4 \AA.

For example, in the most recent opening of OSIRIS for an upgrade in mid-2017, OH emission lines were shifted by 2.8 \AA~from their expected wavelengths using test data taken in the Kn3 35 mas mode. We analyzed and re-derived the wavelength solution for affected data using a standard method involving arc lamp emission line data. The new derived wavelength solution is applicable for data taken after 2017 May 9.

To verify an improvement in the new wavelength solution, we measured offsets between the new wavelength solution and the known position of OH lines in sky frames. Five filters (Kbb, Kn3, Kn1, Jn1, and Hn4) were examined as were two scales (35 and 50 mas). Sky frames in all five filters were used with the 50 mas scale, while two filters, Kbb and Kn3, were also examined with the 35 mas scale.

\begin{figure}
\begin{center}
\includegraphics[width=\columnwidth]{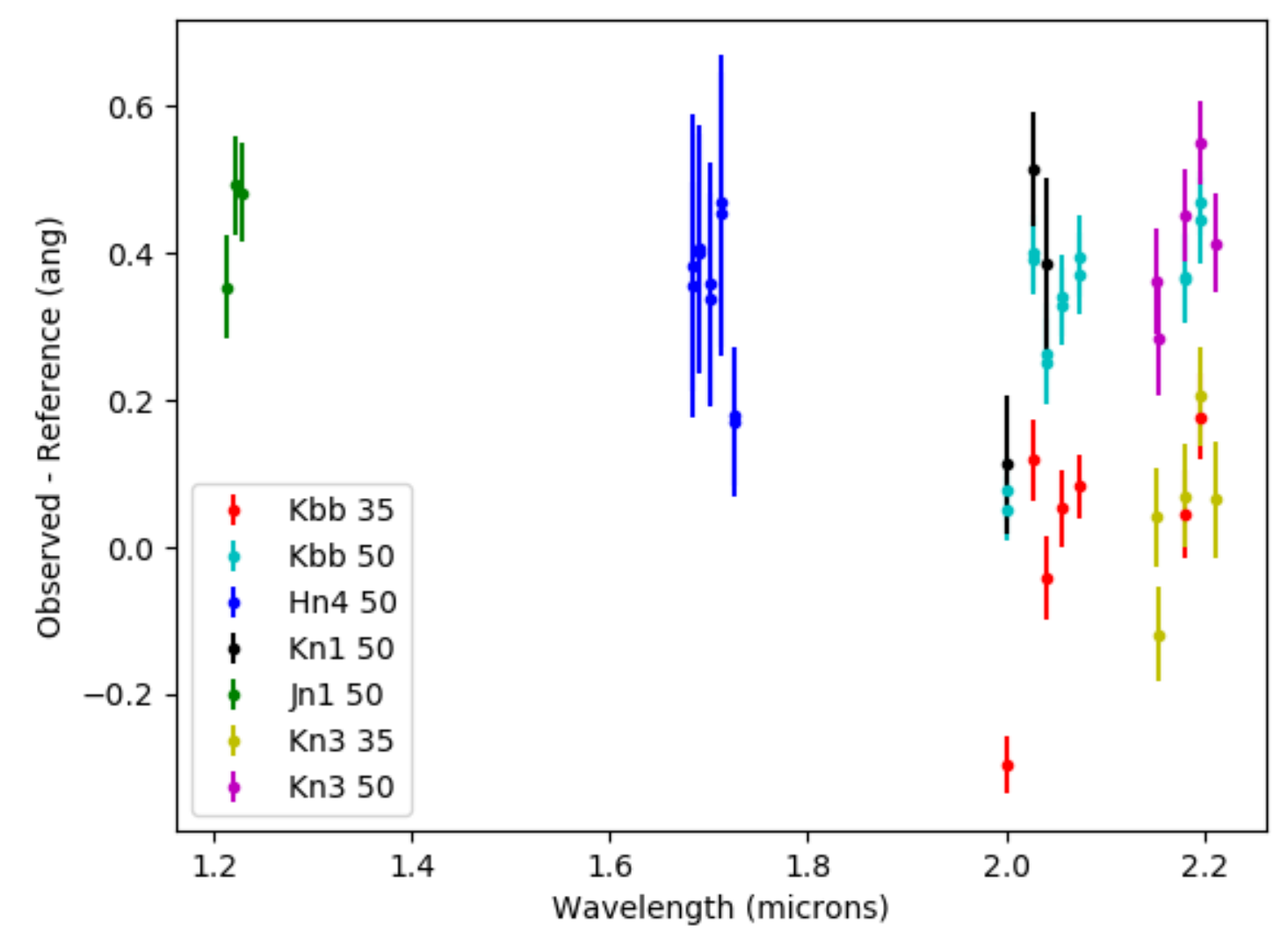}
\caption{Wavelength dependence of the offset of measured OH lines from vacuum wavelength for five broadband filters, after applying the new wavelength solution, valid for data taken 2017 May 9 or later.}
\label{fig:waveoffset}
\end{center}
\end{figure}

For each sky frame, multiple bright OH lines were selected. For each line, the measured wavelength in a given sky frame was taken as the average value for the central 182 lenslets and the error is taken as the standard deviation across the 182 lenslets. The results are shown in figure \ref{fig:waveoffset} as a function of vacuum wavelength and reported in table \ref{tab:waveoffset}. The resulting offsets are suggestive of a scale dependency, with the offsets measured in sky frames taken with the 35 mas scale consistent with zero, while those with a scale of 50 mas have a consistent offset between measured and vacuum wavelength of $\sim$0.4 \AA. This is most evident for the two filters, Kbb and Kn3, tested in both scales.

In general, the wavelength solution has been re-derived whenever the instrument has been opened for major servicing. See table~\ref{tab:wavesol} for a listing of the available wavelength solutions (though note that the DRP will automatically select the appropriate wavelength solution to use based on the given observation's date). However, it is unclear if the wavelength solution has always been verified or re-derived following an instrument opening for minor servicing. Going forward, we recommend verifying the accuracy of the wavelength solution after every instrument opening, even for minor service work, and re-deriving it if necessary.

\begin{deluxetable}{llllllc}
\renewcommand{\arraystretch}{0.9}
\tablecaption{Wavelength Solution Offsets}
\tablewidth{0pt}
\tablehead{
\colhead{Filter} & \colhead{Scale} & \colhead{$\lambda$} & \colhead{Date} & \colhead{Mean$^a$} & \colhead{STD} & \colhead{Num OH}  \\[-2pt]
\colhead{} & \colhead{(mas)} & \colhead{(nm)} & \colhead{} & \colhead{(\AA)} & \colhead{(\AA)} & \colhead{Lines}  \\[-2pt]
}
\startdata
Kbb & 50 & 1965-2381 & 2017-08-12 & 0.327 & 0.127 & 7 \\
Kbb & 50 & 1965-2381 & 2017-08-12 & 0.320 & 0.136 & 7 \\
Kbb & 35 & 1965-2381 & 2017-05-18 & 0.020 & 0.153 & 7 \\
Kn3 & 50 & 2121-2229 & 2017-09-02 & 0.413 & 0.112 & 5 \\
Kn3 & 35 & 2121-2229 & 2017-05-17 & 0.053 & 0.125 & 5 \\
Hn4 & 50 & 1652-1737 & 2017-07-17 & 0.357 & 0.199 & 5 \\
Hn4 & 50 & 1652-1737 & 2017-07-17 & 0.345 & 0.176 & 5 \\
Kn1 & 50 & 1955-2055 & 2017-08-12 & 0.337 & 0.194 & 3 \\
Jn1 & 50 & 1174-1232 & 2017-08-12 & 0.443 & 0.093 & 3 \\
\enddata
\tablenotetext{a}{Mean across all OH lines}
\label{tab:waveoffset}
\end{deluxetable}

\begin{deluxetable}{ccc}
\tablecaption{Available Wavelength Solutions}
\tablehead{
\colhead{Number} & \colhead{Begin Date} & \colhead{End Date} \\
}
\startdata
1 & 2005 Feb 22 & 2006 Feb 22 \\
2 & 2006 Feb 23 & 2009 Oct 04 \\
3 & 2009 Oct 05 & 2012 Jan 03 \\
4 & 2012 Jan 04 & 2012 Nov 09 \\
5 & 2012 Nov 10 & 2015 Dec 31 \\
6 & 2016 Jan 01 & 2017 May 08 \\
7 & 2017 May 09 & 2018 Mar 16 \\
8 & 2018 Mar 16 & present \\
\enddata
\label{tab:wavesol}
\tablenotetext{}{}
\end{deluxetable}

\section{RECOMMENDATIONS}
\label{sec:rec}
Many of the issues we've outlined here for the OSIRIS DRP also have the potential to be of concern for data pipelines for the next generation IFSs on extremely large telescopes, such as IRIS \citep[e.g.][]{2016SPIE.9908E..1WL, 2016SPIE.9913E..4AW} on TMT. These upcoming spectrographs are in their planning phases and both operational daytime and nightime calibrations need to be well-planned, as well as proper maintenance of the data reduction pipeline. We discuss some of the lessons learned here and how they would apply to future IFSs.

We have shown the potential risks of closely packed spectra on the detector with sufficient blending in their PSFs. The deconvolution of raw spectral data is non-trivial, especially when spectra are spaced by a width similar to the FWHM in the spatial direction. In this situation, a 1D spatial rectification algorithm may work with ideal data, but can fall short with real data. Flux may be misassigned (\S~\ref{qsodata}) or cause artifacts in neighboring spaxels (\S~\ref{arclamps}). Ultimately, a 1D estimation of the spectral PSF in the spatial direction may not be an accurate enough representation of real data, particularly for point sources. A 2D estimation of the PSF for every spaxel and wavelength may be necessary to fully resolve this issue. The ability to calibrate the 2D spectral profile of individual spaxels along the detector versus wavelength would greatly benefit post-processing. Calibration routines and analysis methods that explore this 2D spectral extraction ability in next generation IFSs should be further explored.

A similar issue involves the PSF of a point source on the detector in the spectral direction. Sharpening the instrumental PSF allows for higher spectral resolution in science observations, but can induce large-scale patterns. In \S~\ref{sec:spatrip}, we showed that sharpening the spectral PSF to the point of undersampling introduced a spatial rippling effect. This pattern is most evident in single channel maps in reduced data cubes, but when in combination with inadequate spatial rectification, the same pattern can appear in integrated line maps, leading to larger uncertainties in integrated line quantities. The optical design of the spectrograph should be optimized with both spectral width and dispersion that best matches the detector pixel size to perform a high signal-to-noise spectral extraction. This means that the optical engineers need to consider the spectral extraction routines during their layouts.

In \S~\ref{sec:crinrecmat}, we discussed the discovery of improper cosmic ray removal in the rectification matrix calibration files. These noisy spikes in the calibration files propagated through the spectral extraction routine to the reduced science data products. Care should be taken to clean the calibration files as thoroughly as science data before use.

A critical factor in our ability to diagnose and correct issues is the availability of calibration data. We find that data beyond the standard calibration data such as arc-lamp observations and white light scans were necessary in order to assess some of the effects we observed:
\begin{itemize}
\item \textit{Single-column illuminated arc lamps} are needed to evaluate the efficiency of the algorithm that assigns flux to individual spaxels
\item \textit{Single-column illuminated sky observations} are essential for evaluating whether there are relative shifts between the location of the spectra on the detector from the white-light scans compared to on-sky data
\item \textit{A number of sky observations across different filters and plate scales} are necessary to characterize the wavelength solution
\end{itemize}
We recommend that observatories and instrument teams consider the value of additional calibrations to maintain and improve the data products. These calibration data are necessary whenever the instrument hardware is modified and may also be needed whenever the instrument is warmed and opened, even if the major instrument components are unchanged.

\section{CONCLUSIONS}
\label{sec:conc}
We present the first characterization of the OSIRIS DRF using on-sky and calibration data. Our study was begun in response to issues observers had found in their reduced data. As a result of the efforts detailed in this work, the following improvements have been seen in reduced data:
\begin{itemize}
\item We showed that spatial flux assignment artifacts are a result of a pupil mismatch with the 0\farcs1 plate scale, and demonstrated a technique to reduce these artifacts
\item We found that the spectral flux assignment artifacts are due to differences between the PSF of a point source and that of the white light calibration data, and were exacerbated by the narrowness of the PSF in the dispersion direction. Due to this analysis, the dispersion-direction PSF was broadened during a subsequent hardware upgrade in 2016
\item We found evidence of spatial rippling in individual channel maps, due to PSF undersampling. In addition to the spectral flux assignment artifacts, this was also used as evidence to broaden the PSF during the 2016 hardware upgrade
\item We demonstrated that large spikes in reduced data cubes were due to uncorrected cosmic rays in calibration data. We retook the appropriate calibration data when possible and released new rectification matrices without cosmic ray contamination. We also implemented procedures to remove cosmic ray contamination in future rectification matrices
\item We showed that the vast majority of uncorrected ``cosmic rays'' in reduced data cubes were due to bad detector pixels. We released a new bad pixel mask to remove these in reduced data
\item We found inconsistencies in the existing wavelength solution and rederived it to better match on-sky data. We implemented procedures to check the wavelength solution more often going forward
\end{itemize}

Though OSIRIS has been a very productive instrument on both Keck-I and Keck-II for over a decade, its DRP had never been fully characterized using on-sky data. This is almost entirely due to the complexity of the necessary analysis. Our own experiences as members of the OSIRIS Pipeline Working Group showed that maintaining a pipeline of this size and complexity requires a dedicated team of people familiar with the hardware but focused on the software and on the quality of the reduced data. While it's unlikely that observatories, focused on operations and hardware support, can maintain such dedicated software teams, it's vital for the software team to work closely with the observatory instrument team. This eases access to calibration data for the software team, reduces duplicate efforts, and enables knowledge sharing. Finally, though the Keck Observatory Archive\footnote{https://www2.keck.hawaii.edu/koa/public/koa.php} and similar archives offer superb access to years of on-sky data, calibration data is generally not as well archived. However, these data are necessary for this type of work, and it is especially useful to have multiple years of consistent on-sky and calibration data sets to compare.

Our work has implications for teams developing future IFSs. The complexity of IFS raw data outputs means that upcoming IFSs should consider the ease of data reduction from the beginning with their projected spectral overlap, as well as planning for calibration and processing algorithms. Incorporating the lessons learned from the OSIRIS DRP will greatly improve the ease of data reduction and resulting data quality for future IFS instrumentation.

\acknowledgments
The data presented herein were obtained at the W.M. Keck Observatory, which is operated as a scientific partnership among the California Institute of Technology, the University of California and the National Aeronautics and Space Administration. The Observatory was made possible by the generous financial support of the W.M. Keck Foundation. The authors wish to recognize and acknowledge the very significant cultural role and reverence that the summit of Maunakea has always had within the indigenous Hawaiian community.

\bibliographystyle{apj}
\bibliography{references.bib}

\end{document}